\documentclass[aps,prd,10pt,english,preprintnumbers,showpacs,showkeys,
floatfix,nofootinbib,superscriptaddress,fleqn]{revtex4-1} 

\usepackage{bm}
\usepackage{graphicx}
\usepackage{amstext}
\usepackage{amsmath,amsfonts}

\usepackage{hyperref}

\usepackage{babel}

\usepackage{xcolor}

\begin{document}

\title{Flavour structure of the nucleon electromagnetic form factors
  and transverse charge densities in the chiral quark-soliton model}

\author{Ant\'onio Silva}
\email{ajsilva@fe.up.pt}
\affiliation{Departamento de Engenharia F\'isica, Faculdade de
  Engenharia, Universidade do Porto, rua 
  Dr. Roberto Frias, 4200-465 Porto, Portugal}  
\affiliation{  Centro de F\'isica do Porto, Faculdade de Ci\^encias 
da Universidade do Porto, Rua do Campo Alegre 687, 4169-007 Porto,
Portugal    } 

\author{Diana Urbano}
\email{urbano@fe.up.pt}
\affiliation{Departamento de Engenharia F\'isica, Faculdade de
  Engenharia, Universidade do Porto, rua 
  Dr. Roberto Frias, 4200-465 Porto, Portugal} 

\author{Hyun-Chul Kim}
\email{hchkim@inha.ac.kr}
\affiliation{Department of Physics, Inha University, Incheon 402-751, 
  Korea} 
\affiliation{School of Physics, Korea Institute for Advanced Study
  (KIAS), Seoul 130-722, Republic of Korea} 

\date{December, 2017}

\begin{abstract}
We investigate the flavour decomposition of the electromagnetic form
factors of the nucleon, based on the chiral quark-soliton
model ($\chi$QSM) with symmetry-conserving quantisation. We consider
the rotational $1/N_c$ and linear strange-quark mass ($m_s$)
corrections. We discuss the results of the flavour-decomposed 
electromagnetic form factors in comparison with the recent
experimental data. In order to see the effects of the strange quark,
we compare the SU(3) results with those of SU(2). We finally discuss
the transverse charge densities for both unpolarised and polarised
nucleons. The transverse charge density inside a neutron turns out to
be negative in the vicinity of the centre within the SU(3) $\chi$QSM,
which can be explained by the contribution of the strange quark. 
\end{abstract}


\maketitle

\section{Introduction}
Electromagnetic form factors (EMFFs) are the most
fundamental observables that reveal the charge and magnetisation
structures of the nucleon. A series of recent measurements of the
EMFFs has renewed the understanding of the internal structure of the
nucleon and has posed fundamental questions about its nonperturbative 
nature. The results of the ratio of the proton EMFFs, $\mu_p
G_{E}^p/G_{M}^p$ with the proton magnetic moment $\mu_p$, obtained by
measuring the transverse and longitudinal recoil proton
polarisations~\cite{Jones:1999rz,
  Gayou:2001qt,Gayou:2001qd,Punjabi:2005wq,  
Puckett:2010ac, Bernauer:2010wm, Ron:2011rd,
  Zhan:2011ji}, were found to decrease almost   linearly with $Q^2$
above $1\,(\mathrm{GeV}/c)^2$.   These results were in
conflict with most of the previous measurements of the proton EMFFs from
unpolarised electron-proton cross sections based on the Rosenbluth
separation method.
These new experimental results have triggered subsequent theoretical
and experimental works (see, for example, recent
reviews~\cite{HydeWright:2004gh, Arrington:2006zm, Perdrisat:2006hj,
  Vanderhaeghen:2010nd, Arrington:2011kb}). This discrepancy is
partially explained by the effects of two-photon exchange, which
affect unpolarised electron-proton scattering at higher $Q^2$ but have
less influence on the polarisation measurements~\cite{Guichon:2003qm,
  Blunden:2003sp, Arrington:2004ae, Arrington:2007ux, Carlson:2007sp,
  Arrington:2011dn}. Moreover, the new experimental results of the
proton EMFFs in a wider range of $Q^2$ provided a whole new
perspective on the internal quark-gluon structure of the
nucleon. Perturbative quantum chromodynamics (pQCD) with factorisation
schemes~\cite{Brodsky:1974vy} predicts the different scalings of the
Dirac and Pauli FFs, $F_1^p$ and $F_2^p$: $F_1^p$ falls off as $1/Q^4$
while $F_2^p$ decreases as $1/Q^6$, so that $Q^2 F_2^p/F_1^p$ becomes
flat at large $Q^2$. However, the experimental data show that the
ratio $Q^2 F_2^p/F_1^p$ increases with $Q^2$ but $Q F_2^p/F_1^p$
becomes flat starting around $2\,\mathrm{GeV}^2$. 
A similar discrepancy between the experimental data and pQCD was also
found in the $\gamma\gamma^*\to\pi$ transition form
factor~\cite{Aubert:2009mc,Uehara:2012ag} even for higher $Q^2$.  
It implies that  it is far more important to consider effects from
nonperturbative physics than those from perturbative QCD in lower
$Q^2$ region.    

Assuming isospin and charge symmetries, neglecting the strangeness in
the nucleon, and using both the experimental data for the proton and
neutron EMFFs, Cates \textit{et al.}~\cite{Cates:2011pz} have
extracted the up and down EMFFs and have obtained 
remarkable results: the $Q^2$ dependence of the up- and down-quark
Dirac ($F_1$) and Pauli ($F_2$) form factors (FFs) are considerably
different from each other. The down-quark Dirac and Pauli FFs are
roughly proportional to $1/Q^4$ but those of the up-quark fall off
more gradually. Moreover, while the ratios $\kappa_u^{-1} F_2^u/F_1^u$
and $\kappa_d^{-1} F_2^d/F_1^d$ ($\kappa$ is the anomalous magnetic
moment) are relatively constant above $Q^2\sim 1\,\mathrm{GeV}^2$,
they show a complicated behavior for lower $Q^2$ regions. Qattan and 
Arrington~\cite{Qattan:2012zf,Qattan:2015qxa} elaborated on the analysis of
Ref.~\cite{Cates:2011pz}, taking into account explicitly the effects
of two-photon exchange and uncertainties on the proton form factor and 
the neutron magnetic FFs.  They found that the ratio of the
up-quark EMFFs ($G_E^u/G_M^u$) has a roughly linear drop-off, while
that of the down-quark EMFFs ($G_E^d/G_M^d$) showed a completely
different dependence on $Q^2$. As a result, the flavour-decomposed
FFs behave in a different way from the proton EMFFs.
  Diehl and Kroll~\cite{Diehl:2013xca} critically analyzed 
experimental data in  order to study several hadron properties and also 
obtained a separation of the light quark contributions to form factors.
The flavour contributions to the EMFFs of the nucleon and the related charge and 
magnetization densities had already been the subject of interest prior to 
the phenomenological analysis of \cite{Cates:2011pz} and of \cite{Qattan:2012zf,
Diehl:2013xca}: Ref.~\cite{Cloet:2008re} used a framework based on the Faddeev equation
with dressed quarks to obtain the flavor contributions to the Dirac, Pauli and 
Sachs form factors, including the associated radii, 
while \cite{Crawford:2010gv} used a vector dominance model. 
These studies, as well as experimental results \cite{Riordan:2010id}, 
pointed out the nontrivial behaviour of these contributions as revealed further 
by the analysis of \cite{Cates:2011pz,Qattan:2012zf,Diehl:2013xca}. 
Several theoretical studies of these contributions have since been performed:
Ref.~\cite{Eichmann:2011vu} further developed the covariant Faddeev framework, based on 
the Dyson-Schwinger equations of QCD,
~\cite{Rohrmoser:2011tw,Rohrmoser:2017zpk}  
employed a Goldstone-boson-exchange relativistic constituent quark model, 
~\cite{Cloet:2012cy} extended the quark-diquark model to include a pion cloud.
In Ref.~\cite{GonzalezHernandez:2012jv} the flavour contributions to the EMFFs 
were obtained by computing the generalized parton distributions in a reggeized 
diquark model and in \cite{Chakrabarti:2013dda} from generalized parton 
distributions obtained in a quark model in AdS/QCD. The AdS/QCD correspondence 
has been the basis for similar studies within different approaches: 
in a light-front quark model in a soft-wall model~\cite{Mondal:2015uha}; 
or a hard-wall model~\cite{Mondal:2016xpk}; also via parameterization 
approaches,~\cite{Sharma:2016cnf} and \cite{Nikkhoo:2015jzi}. 
Ref.~\cite{Brodsky:2016uln}  used the light-front holographic QCD framework 
including higher Fock components and 
Ref.~\cite{Obukhovsky:2014xja} a relativistic light-front model. 
The flavour contributions may equally be displayed through the 
transverse charge and magnetization densities, as one may find in 
Ref.s~\cite{Chakrabarti:2014dna,Sharma:2014voa},
which employed a soft-wall model of AdS/QCD, and also in some 
of the aforementioned studies.        

In this context, we investigated the flavour structure of the
nucleon EMFFs within the 
framework of the self-consistent SU(2) and SU(3) chiral quark-soliton
models ($\chi$QSMs)~\cite{Diakonov:1987ty,
  Wakamatsu:1990ud,Blotz:1992pw}. The $\chi$QSM has described
successfully various observables of the baryon octet and decuplet (For
reviews, see~\cite{Alkofer:1994ph, Christov:1995vm,  Weigel:2008zz, 
  Goeke:2001tz}). In particular, the $Q^2$ dependence of almost all
form factors is well reproduced within the $\chi$QSM, so that the 
strange-quark EMFFs~\cite{Goeke:2006gi} and the parity-violating (PV) 
asymmetries of polarised electron-proton
scattering~\cite{Silva:2005qm}, which require nine different FFs (six
EMFFs and three axial-vector FFs) with the same set of parameters, are
in good agreement with experimental data.
Thus, it is worthwhile to examine the flavour
structure of the nucleon EMFFs in detail. As mentioned, the nucleon
EMFFs were already studied in the SU(3)
$\chi$QSM~\cite{Kim:1995mr}. However, Prasza\l owicz \textit{et
  al.}~\cite{Praszalowicz:1998jm} pointed out that the
Gell-Mann-Nishijima relation was not exactly fulfilled in the initial
version of the $\chi$QSM and proposed the symmetry-conserving
quantisation that makes the Gell-Mann-Nishijima relation well
satisfied. We want to emphasize that the $\chi$QSM is a reasonable
framework to investigate properties of the lowest-lying SU(3)
baryons. Witten originally proposed in his seminal
papers~\cite{Witten:1979kh, Witten:1983} that the 
lowest-lying light baryons may be regarded as bound states of $N_c$
\textit{valence} quarks in a meson mean field. In the 
limit of the large number of colours ($N_c$), the lowest-lying SU(3) 
baryons constitute $N_c$ \textit{valence} quarks that bring about an
effective pion mean field or the vacuum polarization. The value of
$N_c$ will be taken to be three at the final stage of the
computation such that we are able to compare the present results with
the experimental data. Recently, this mean-field approach or the
$\chi$QSM described successfully properties of singly heavy 
baryons~\cite{Yang:2016qdz, Kim:2017jpx, Kim:2017khv}.   

In this work, we present the results of the flavour-decomposed up-  and
down-quark EMFFs based on the SU(3) $\chi$QSM with symmetry-conserving
quantisation employed. We first show the Dirac and Pauli FFs of the nucleon
and then examine the $Q^2$ dependence of the up- and down-quark Dirac
and Pauli FFs. The ratio of the flavour-decomposed Dirac and Pauli FFs
will be discussed, compared with the recent experimental
data~\cite{Qattan:2012zf}. We also reexamine the results of the 
strange EMFFs, since there are new experimental data from
PV polarised electron-nucleon scattering. In particular, the G0
collaboration recently measured the paritiy-violating asymmetries in
the backward angle~\cite{Androic:2009aa}, which was first predicted in
Ref.~\cite{Silva:2005qm}. In addition to the flavour decomposed EMFFs
of the nucleon, we also investigate the charge and magnetisation
densities of the quark in a nucleon in the transverse plane. Together
with the new experimental data for the nucleon EMFFs, the nucleon GPDs
cast light on the concept of nucleon FFs~\cite{Ji:1998pc,
  Goeke:2001tz, Diehl:2003ny, Belitsky:2005qn}. 

The present work is sketched as follows. In Section 2, we
briefly review the general formalism of the EMFFs of the nucleon and
its flavour decomposition and describe how to compute the EMFFs of the
nucleon within the framework of the SU(2) and SU(3) $\chi$QSMs. In the
following sections we present the results and discuss their physical
implications in the light of the recent experimental data: for Sachs
FFs in Section 3 and for the Dirac and Pauli FFs in Section 4. In
section 5 we also present the model results for the transverse charge
and magnetic distributions of the quark inside both unpolarised and 
transversely polarised nucleons. The final Section is devoted to the
summary and the conclusions.     

\section{Electromagnetic Form factors and the $\chi$QSM}
The matrix element of a flavour vector current between the two
nucleon states is expressed in terms of the flavour Dirac and Pauli
FFs 
\begin{eqnarray}
\lefteqn{\langle N (p',\,s')|J_\mu^{\chi} (0)|
N(p,\,s)\rangle } \hspace{1cm}\cr
&& =  \bar{u}_{N}(p',\,s') 
\left[ \gamma _{\mu }F_{1}^\chi(q^{2})+i\sigma _{\mu \nu }
\frac{q^{\nu }}{2M_{N}}F_{2}^\chi (q^{2})\right] u_{N}(p,\,s),
\label{eq:ff1}  
\end{eqnarray}
where $J_\mu ^{\chi} (0)$ represents the flavour vector current defined
as 
\begin{equation}
  \label{eq:flavor_current}
J_\mu^{\chi}(0) \;=\; \bar{\psi}(0) \lambda^{\chi} \gamma_\mu
\psi(0).   
\end{equation}
$\chi$ denotes the flavour index, i.e. $\chi=0,\,3,\,8$ for the flavour
decomposition. Here, one has to bear in mind that $\lambda^0$ is
considered to be a unity flavour matrix. Thus, the normalisation
$\{\lambda^a,\,\lambda^b\}=2\delta^{ab}$ for
$\lambda^\chi$ applies only to the Gell-Mann matrices with $\chi=3$
and $\chi=8$. The Dirac spinor $u_N(p,\,s)$  
applies to the nucleon with mass $M_N$, momentum $p$ and the
third component of its spin $s$. The square of the four momentum 
transfer is denoted by  $q^{2}=-Q^{2}$, with $Q^{2}>0$. 
The flavour Dirac and Pauli FFs can be combined to give the Sachs FFs: 
\begin{eqnarray}
G^{\chi}_{E}(Q^{2}) & = & F_1^{\chi}(Q^{2})
-\frac{Q^{2}}{4M^{2}_{N}} F_2^{\chi}(Q^{2})\nonumber \\
G_{M}^{\chi}(Q^{2}) & = & F_{1}^N(Q^{2})+F_2^{\chi}(Q^{2}).
\label{eq:DiracPauli2Sachs}
\end{eqnarray}
In the Breit frame, $G_{E}^\chi(Q^{2})$ and
$G_M^{\chi}(Q^{2})$ are related to the time and space components of the
flavour vector current, respectively:  
\begin{eqnarray}
G^{\chi}_{E}(Q^{2})&=&\langle N'(p')|\bar{\psi}(0)\gamma_0\lambda^\chi 
\psi(0)|N(p)\rangle \nonumber \\ 
G^{\chi}_{M}(Q^{2}) & = &i M_N \epsilon_{ilk} \frac{q_l}{6q^2} 
{\rm tr}\left(\langle p',\lambda'| \bar{\psi}
  (0)\gamma_{i}\lambda^\chi 
\psi(0)  |p,\lambda\rangle \sigma_k \right).
\label{Eq:gm} 
\end{eqnarray}
 where $\sigma _{j}$ are the Pauli spin matrices. 
The $|\lambda \rangle$ is the corresponding spin state of the 
nucleon.  

In SU(3) flavour the nucleon EMFFs are expressed in terms of the
triplet and octet vector form factors  
\begin{equation}
G_{E,M}^N (Q^2) \;=\; \frac12\left(G_{E,M}^3 + \frac1{\sqrt{3}}
G_{E,M}^8\right),
\label{eq:Gsu3}   
\end{equation}
while in flavour SU(2) they are written as
\begin{equation}
G_{E,M}^N (Q^2) \;=\; \frac12\left(\frac13 G_{E,M}^0 + 
G_{E,M}^3\right).   
\label{eq:Gsu2}   
\end{equation}
Although the same notation is used for the form factors, it will
always follow from the context which flavour case is being addressed.  

The matrix elements given in Eq.~(\ref{Eq:gm}) can be evaluated both
in the SU(2) and SU(3) flavour $\chi$QSMs. The model starts from the
following low-energy effective partition function in Euclidean space 
\begin{eqnarray}
\mathcal{Z}_{\mathrm{\chi QSM}} &=&
\int\mathcal{D}\psi\mathcal{D}\psi^{\dagger}\mathcal{D}U \exp
\left[-\int   d^{4}x\Psi^{\dagger}iD(U)\Psi\right] \cr
&=& \int\mathcal{D}U \exp(-S_{\mathrm{eff}}[U])\,, 
\label{eq:part}    
\end{eqnarray}
where $\psi$ and $U$ denote the quark and pseudo-Goldstone boson
fields, respectively. After integrating over the quark fields, 
the effective chiral action $S_{\mathrm{eff}}$ is given by 
\begin{equation}
S_{\mathrm{eff}}(U) \;=\; -N_{c}\mathrm{Tr}\ln iD(U)\,,
\label{eq:echl}
\end{equation}
where $\mathrm{Tr}$ represents the functional trace and $N_c$ the number 
of colours.

The Dirac  $D(U)=\gamma_{4}(i\rlap{/}{\partial} - \hat{m} -
MU^{\gamma_{5}})$  operator, depending on the flavour space, is given
by   
\begin{eqnarray}
D_\mathrm{SU(2)}(U) &=&   -i\partial_{4} + h(U)  \cr
D_{\rm SU(3)}(U)    &=&   -i\partial_{4} + h(U) - \gamma_{4}\delta m
\label{eq:Dirac}  
\end{eqnarray}
since, as isospin symmetry is assumed in this work,
$\hat{m}=\mathrm{diag}(\overline{m},\,\overline{m}) =
\overline{m}\mathbf{1}_2$ in SU(2) and
$\hat{m}=\mathrm{diag}(\overline{m},\, \overline{m}, \,
m_{\mathrm{s}})=\overline{m}\mathbf{1}_3+\delta m$ in SU(3), where 
\begin{equation}
\delta m  \;=\; \frac{-\overline{m} + m_{s}}{3}\mathbf{1}_3 +
\frac{\overline{m} - m_{s}}{\sqrt{3}} \lambda^{8} =
M_{1} \mathbf{1}_3 + M_{8}  \lambda^{8}\,.
\label{eq:deltam}
\end{equation}
The mass term $\delta m$ containing
the strange current quark mass $m_s$ will be treated as a
perturbation.

The single-quark Hamiltonian $h(U)$ is expressed as 
\begin{equation}
h(U) \;=\;
i\gamma_{4}\gamma_{i}\partial_{i}-\gamma_{4}MU^{\gamma_{5}} -
\gamma_{4} \overline{m}\, ,
\label{eq:diracham}  
\end{equation}
where $U^{\gamma_5}$ stands for the chiral field for which we assume
Witten's embedding of the SU(2) soliton
into SU(3)  
\begin{equation}
U^{\gamma_{5}}_{\mathrm{SU(3)}} \;=\; \left(\begin{array}{lr}
U_{\mathrm{SU(2)}}^{\gamma_{5}} & 0\\
0 & 1
\end{array}\right)
\label{eq:embed}
\end{equation}
with the SU(2) pion field $\pi^i$ as 
\begin{equation}
U_{\mathrm{SU(2)}}^{\gamma_{5}} \;=\; \exp(i\gamma^{5}\tau^i \pi^i)
\;=\; \frac{1+\gamma^{5}}{2}U_{\mathrm{SU(2)}} +
\frac{1-\gamma^{5}}{2}U_{\mathrm{SU(2)}}^{\dagger}. 
\label{eq:u2}
\end{equation}
The integration over the pion field $U$ in Eq.~(\ref{eq:part}) can be
performed by the saddle-point approximation in the large $N_c$ limit
due to the $N_{c}$ factor in Eq. (\ref{eq:echl}).  The SU(2) pion
field $U$ is written as the most symmetric hedgehog form  
\begin{equation}
U_{\mathrm{SU(2)}} \;=\; \exp[i\gamma_{5}\hat{\bm n}\cdot{\bm
  \tau}P(r)]\,  ,
\label{eq:hedgehog}  
\end{equation}
where $P(r)$ is the radial profile function of the soliton.

The $\chi$QSM nucleon state $|N(p,\,s)\rangle$ used in the
computation of Eqs.~(\ref{eq:ff1}) and (\ref{Eq:gm}) is defined in
terms of an Ioffe-type current consisting of $N_c$ quarks:  
\begin{equation}
|N(p,\,s)\rangle \;=\; 
\lim_{x_{4}\rightarrow-\infty}\,\frac{1}{\sqrt{\mathcal{Z}}}\,  
e^{ip_{4}x_{4}}\,\int d^{3}{\bm x}\, e^{i\,{\bm p}\cdot{\bm x}}\,
J_{N}^{\dagger}(x)\,|0\rangle
\label{eq:Ioffe}  
\end{equation}
with the Ioffe-type nucleon current $J_N$ defined as 
\begin{equation}
J_{N}(x) \;=\; \frac{1}{N_{c}!}\,\Gamma_{N}^{b_{1}\cdots
  b_{N_{c}}}\,\varepsilon^{\beta_{1}\cdots\beta_{N_{c}}}\,\psi_{\beta_{1}b_{1}}(x)
\cdots\psi_{\beta_{N_{c}}b_{N_{c}}}(x)\,.
\end{equation}
Here, the matrix $\Gamma_N^{b_{1}...b_{N_{c}}}$ carries the
hypercharge $Y$, isospin $I,I_{3}$ and spin $s,s_{3}$ quantum numbers
of the baryon and the $b_{i}$ and $\beta_{i}$ denote the spin-flavour
and colour indices, respectively. 

After minimizing the action in Eq. (\ref{eq:echl}), we derive an
equation of motion which is solved self-consistently with respect to
the function $P(r)$ in Eq.~(\ref{eq:hedgehog}). The corresponding
unique solution $U_c$ is called the classical chiral soliton. 
The next step consists in quantising the classical soliton. This can
be achieved by quantising the rotational and translational zero-modes
of the soliton. The rotations and translations of the soliton are 
implemented by  
\begin{equation} 
U({\bm x},t)=A(t)U_{c}({\bm x}-{\bm
    z}(t))A^{\dagger}(t)\,,
\end{equation} 
where $A(t)$ denotes a time-dependent SU(3) matrix, related to the  
orientation of the soliton in coordinate and flavour spaces, and ${\bm
  z}(t)$ stands for the time-dependent translation of the centre of
mass of the soliton in coordinate space. The rotational velocity of
the soliton $\Omega(t)$ is defined as  
\begin{equation}
\Omega=\frac{1}{i}A^{\dagger}\dot{A} =
\frac{1}{2i}\textrm{Tr}(A^{\dagger}\dot{A}\lambda^{\alpha}) 
\lambda^{\alpha}=\frac{1}{2}\Omega_{\alpha}\lambda^{\alpha}.
\end{equation}  
Treating $\Omega(t)$ and $\delta m$ perturbatively with
slowly rotating soliton and small $\delta m$ considered, we find the
collective Hamiltonian, i.e, the Hamiltonian in the collective 
coordinates of position of the centre of mass and the orientation of
the soliton, which is given explicitly as 
\begin{equation}
H_\mathrm{coll}^\mathrm{SU(2)}  =
M_{c}^\mathrm{SU(2)}+\frac{1}{2I_{1}^{\mbox{\tiny
      SU(2)}}}\sum_{i=1}^{3}J_{i}J_{i}   
\end{equation}
in SU(2) and as 
\begin{eqnarray}
H_\mathrm{coll}^\mathrm{SU(3)} & = & H_{\mathrm{sym}}+H_{\mathrm{sb}}\\ 
H_{\mathrm{sym}} & = & M_{c}+\frac{1}{2I_{1}}\sum_{i=1}^{3}J_{i}J_{i}
+ \frac{1}{2I_{2}}\sum_{a=4}^{7}J_{a}J_{a},\\
H_{\mathrm{sb}} & = & \frac{1}{\overline{m}}M_{1}\Sigma_{SU(2)} +
\alpha D_{88}^{(8)}(A) +\beta
Y+\frac{\gamma}{\sqrt{3}}D_{8i}^{(8)}(A)J_{i}\,.
\label{eq:Ham}
\end{eqnarray}
in SU(3). The $M_c$ is the classical mass of the state, the parameters
$I$ are inertia parameters, $Y$ is the hypercharge, $\Sigma_{SU(2)}$
is the pion-nucleon sigma term, the $J$s are the angular momentum 
operators and $D^{(8)}$ are the SU(3) Wigner $D$ functions. It is
obvious that the strange quark in flavour SU(3) leads to a more
involved analysis, particularly to the symmetry breaking
contributions.       

Within the collective quantisation procedure the nucleon states 
given in Eq.~(\ref{eq:Ioffe}) will be mapped to collective rotational functions
carrying the state quantum numbers.   
In flavour SU(2) these functions are the eigenfunctions of the SU(2)  
symmetrical Hamiltonian, i.e., the Wigner $D$ functions given as 
\begin{eqnarray}
\Psi_{JJ_3TT_3}(A)&=&\langle A \vert N (JJ_3;\,TT_3) \rangle \cr
&=& (-1)^{T+T_3}\sqrt{2T+1}D^ {T=J}_{-T_3,J_3}(A).
\label{eq:psiSU2} 
\end{eqnarray}
In flavour SU(3) the eigenfunctions of the SU(3) symmetric part 
of the Hamiltonian turn out to be the SU(3) Wigner $D$ functions 
\begin{eqnarray}
\Psi^n_{Y;JJ_3;TT_3}(A)&=&\langle A \vert N (Y;\,JJ_3;\,TT_3) \rangle \cr
&=& \sqrt{\dim n}(-1)^{-1/2+J_3}D^{(n)*}_{T,T_3,Y;J,J_3,-1}(A).
\label{eq:psiSU3} 
\end{eqnarray}
On the contrary to the SU(2) case, the nucleon state is no longer a
pure octet state but is a mixed state with those in higher
representations arising from flavour SU(3) symmetry breaking, i.e.  
\begin{eqnarray}
\lefteqn{ |N (Y;JJ_3;TT_3)\rangle \;=\; |8_{1/2}\:(Y;JJ_3;TT_3)\rangle
} \hspace{7mm} \cr 
&&  + c_{\overline{10}}\sqrt{5}|
\overline{10}_{1/2}\:(Y;JJ_3;TT_3)\rangle  + c_{27}\sqrt{6}|
27_{1/2}\:(Y;JJ_3;TT_3)\rangle,  
\label{eq:Nwave}
\end{eqnarray}
where $c_{\overline{10}}$ and $c_{27}$ denote the mixing parameters. These 
parameters, as well as the $\alpha$, $\beta$ and $\gamma$ in
Eq.~(\ref{eq:Ham}), may be found in
Refs.~\cite{Blotz:1992pw,Christov:1995vm}.  

A detailed formalism for the zero-mode quantisation can be found in
Refs.~\cite{Blotz:1992pw,Christov:1995vm}.  In addiction,
Ref.~\cite{Kim:1995mr} offers a detailed description as to how the
form factors can be obtained numerically.  
We briefly summarise it here before we discuss the numerical results. 
The parameters existing in the model are the constituent quark mass
$M$, the current quark mass $\overline{m}$, the strange current quark
mass $m_{\mathrm{s}}$, and the cutoff mass $\Lambda$ of the
proper-time regularisation. However, not all of them are free
parameters but can be fixed in the mesonic sector without any
ambiguity. In fact, this is a merit of the $\chi$QSM in which mesons
and baryons can be treated on an equal footing. For a given $M$ the
regularisation cut-off parameter $\Lambda$ and the current quark mass
$\overline{m}$ in the Lagrangian are fixed to the pion decay constant
$f_{\pi}=93$ MeV and the physical pion mass $m_{\pi}=140$\, MeV,
respectively. The strange current quark mass is taken to be
$m_{\mathrm{s}}=180\,\textrm{MeV}$ which approximately reproduces the
kaon mass. Though the constituent quark mass $M$ can be regarded as a
free parameter, it is also more or less fixed. The experimental proton
electric charge radius is best reproduced in the $\chi$QSM with the
constituent quark mass $M=420$ MeV.  Moreover, the value of 420 MeV is
known to yield the best fit to many baryonic
observables~\cite{Christov:1995vm}. Thus, all the numerical results in
the present work are obtained with this value of $M$. 

All the results presented in the following were computed completely
within the model, in the same level of approximation, to keep 
consistency. In particular, magnetisation observables are presented 
not in terms of the physical nuclear magneton but, instead, in terms
of the model nuclear magneton, i.e. defined as the model value for the 
nucleon mass, which, at this level of approximation used in this work,
is     
\begin{equation}
M_N=1250\;\textrm{MeV}.
\label{eq:M}
\end{equation}
We want to mention that the ratio between the model nuclear magneton
and the physical one is the same as that between the value of $M_N$ in
Eq.(\ref{eq:M}) and the physical nucleon mass.

To address the properties of the baryon octet implies immediately 
flavour structures of the SU(3) baryons. However, it indicates
simultaneously the question as to how accurate is the $\chi$QSM
description of the strangeness content of the nucleon and its
implications to  the EMFF. Such a question could easily be answered if
one had precise experimental data on the strange EMFF. The present 
study may give some clues to the answer for that question in the light
of the recent phenomenological data~\cite{Cates:2011pz,Qattan:2012zf}.  

\section{Sachs form factors}
The Sachs EM form factors~\cite{Ernst:1960zza,Sachs:1962zzc} are the  
most common form to encompass information about the
electromagnetic structure of the nucleon.  
On the one hand, these form factors make it possible to express the
cross section for elastic electron-proton scattering in the one-photon
exchange approximation, without mixed terms ($G_EG_M$) in a form
suitable for the separation of the electric and magnetic form
factors.  That is not the case when the cross section is expressed in
terms of the Dirac and Pauli form factors~(\ref{eq:ff1}), where the
mixed terms ($F_1F_2$) occur. Even with the more recent polarisation
transfer methods~\cite{Arnold:1980zj}, the measured ratio between the
longitudinal and transverse polarisation components is expressed in
terms of the Sachs form factors ratio $\mu G_E/G_M$.   

On the other hand, the Sachs form factors have a merit that in the 
Breit frame they may be apparently interpreted as the Fourier
transform of the charge and magnetisation distributions inside a
nucleon. It comes from the fact that in the Breit frame the proton
does not exchange energy with the virtual photon with momentum
$(0,\,{\bm q})$. At a specific space-like $Q^2=-{\bm q}^2<0$  
invariant momentum transfer, the time and space components of the
electromagnetic current, associated with the electric and magnetic
form factors respectively, resemble the classical non-relativistic
current density. Hence the Sachs EM form factors are directly related
to the charge and magnetisation distributions by the Fourier 
transform. However, these relations are supposedly non-relativistic in 
nature due to the $Q^2$ dependence of the Breit-frame. 
Both the preceding features of the Sachs form factors are currently
under scrutiny, as mentioned in Introduction. Discrepancies in the
experimental results from the elastic $ep$ cross section and
polarisation transfer studies called for the inclusion of new aspects
of elastic electron proton scattering, such as the 
two-photon exchange~\cite{Carlson:2007sp}. The connection between form    
factors and densities, even apart from the non-relativistic
limitation, has also been revised on general
grounds~\cite{Miller:2007uy,Miller:2010nz}.         

\begin{figure}[ht]
\centering{\includegraphics[width=0.49\textwidth]{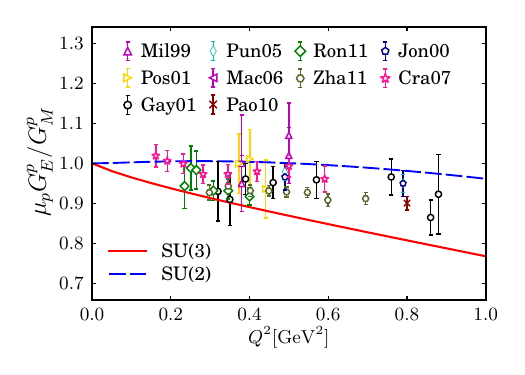}
\includegraphics[width=0.49\textwidth]{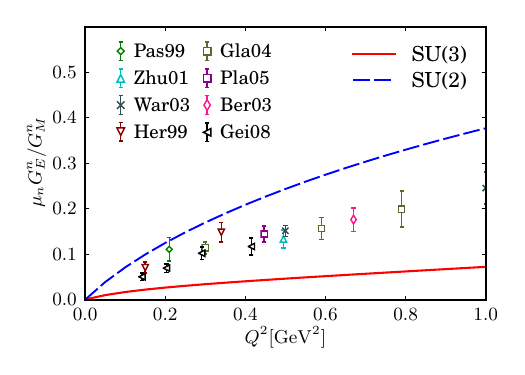}}
\caption{(Color online) The ratio of the proton magnetic FF to the
  electric FF: $\mu G_E/G_M$ in the left panel.  
The experimental data are taken from 
Mil99 \cite{Milbrath:1997de}, 
Pos01 \cite{Pospischil:2001pp}, 
Gay01 \cite{Gayou:2001qt}, 
Mac06 \cite{MacLachlan:2006vw}, 
Pao10 \cite{Paolone:2010qc}, 
Ron11 \cite{Ron:2011rd}, 
Zha11 \cite{Zhan:2011ji}, 
Jon06 \cite{Jones:2006kf}, 
Jon00 \cite{Jones:1999rz}, 
Cra07 \cite{Crawford:2006rz}.
The neutron $\mu G_E/G_M$ ratio in the right panel compared to the
data from recent experiments:  
Pas99 \cite{Passchier:1999cj}, 
Zhu01 \cite{Zhu:2001md}, 
War03 \cite{Warren:2003ma}, 
Gei08 \cite{Geis:2008aa}, 
Her99 \cite{Herberg:1999ud}, 
Gla04 \cite{Glazier:2004ny}, 
Pla05 \cite{Plaster:2005cx}, 
Ber03 \cite{Bermuth:2003qh}. 
The solid curve depicts the result from the SU(3) $\chi$QSM whereas
the dashed one does that from the SU(2) model.
}
\label{fig:1} 
\end{figure} 
In the left panel of Fig.~\ref{fig:1}, the results of the ratio of the
proton magnetic FF to the electric FF are depicted in comparison with
the experimental data from the recoil polarisation experiments  
$p(\vec{e},e'\vec{p})$~\cite{Jones:1999rz, Gayou:2001qt,
  Puckett:2010ac,Ron:2011rd,  Zhan:2011ji,
  Milbrath:1997de,Pospischil:2001pp, MacLachlan:2006vw,Paolone:2010qc,
  Puckett:2011xg}   
and the experiments with a polarised target
$\vec{p}(\vec{e},e'\vec{p})$~\cite{Jones:2006kf,Crawford:2006rz}.
The SU(2) results can describe the general tendency of the data very 
well, whereas those of SU(3) seem slightly underestimated, as $Q^2$ 
increase. The right panel of Fig.~\ref{fig:1} plots the results for
the ratio $\mu_nG_E^n/G_M^n$, compared with the experimental data
taken from $\vec{d}(\vec{e},en)p$~\cite{Passchier:1999cj, 
Zhu:2001md,Warren:2003ma,Geis:2008aa} and from
$d(\vec{e},e'\vec{n})p$~\cite{Herberg:1999ud,Glazier:2004ny,
Plaster:2005cx} and $^3\vec{\mathrm{He}}(\vec{e},e'n)$
scatterings~\cite{Bermuth:2003qh}.
We observe that the experimental data lie between the SU(2) and SU(3)
results. The general tendency of the present results are in line with
the experimental data: $\mu_pG_E^p/G_M^p$ falls off slowly as $Q^2$ 
increases, while $\mu_nG_E^n/G_M^n$ increases systematically as 
a function of $Q^2$.  As shown in the right panel of Fig~\ref{fig:1},
the SU(3) results for the neutron are rather different from those in
SU(2), the reason stemming, at least partially, from the strange quark
contribution to the neutron electric FF.  Because of the embedding of
the SU(2) soliton into SU(3) as shown in Eq.(\ref{eq:embed}), the
contribution of the strange quark has the same asymptotic behavior of
the nonstrange quarks. The effects due to different asymptotic tails
were discussed in Ref.~\cite{Silva:2001st} in the context of
the strange vector FFs of the nucleon. Thus, in a sense, a true answer
may be found between the SU(2) and the SU(3) results. 

In order to decompose the proton EMFFs into the flavour ones, we need 
to compute the singlet vector form factors of the proton. Then, we are
able to express the flavour-decomposed EMFFs of the proton in terms of 
the singlet, triplet, and octet FFs of the proton: 
\begin{eqnarray}
G_{E,M}^u (Q^2) &=& \frac12 \left(\frac23 G_{E,M}^{(0)} (Q^2) +
  G_{E,M}^{(3)} (Q^2) + \frac1{\sqrt{3}}   G_{E,M}^{(8)}
  (Q^2)\right),\cr 
G_{E,M}^{d}(Q^2) &=& \frac12 \left(\frac23 G_{E,M}^{(0)} (Q^2) - 
G_{E,M}^{(3)} (Q^2) + \frac1{\sqrt{3}} G_{E,M}^{(8)}
  (Q^2)\right),\cr  
G_{E,M}^{s}(Q^2) &=& \frac13 \left(G_{E,M}^{(0)} (Q^2) -
\sqrt{3} G_{E,M}^{(8)} (Q^2)\right),
  \label{eq:flavor_decompose}
\end{eqnarray}
where we have suppressed the corresponding quark charge. The
normalisations at $Q^2=0$ for the proton obey $G_E^u (0)=2$, $G_E^d
(0)=1$ and $G_E^s (0)=0$. The flavour-decomposed magnetic moments are
listed in Table~\ref{tab:1} in unit of the model nuclear magneton, 
{\em i.e.} defined with the model nucleon mass.  
\begin{table}[h] \begin{center}
\begin{tabular}{cccc} \hline
   & $\mu_u$ & $\mu_d$ & $\mu_s$ \\
 \hline \hline
SU(3)  & $3.22$ & $-0.73$ & $0.10$ \\
SU(2)  & $3.46$ & $-0.95$ &  \\
\cite{Qattan:2012zf}&  $3.67$ & $-1.03$ &  \\
\hline
\end{tabular}
\end{center}
\caption{The flavour-decomposed magnetic moments are defined as
  $\mu_q=G_M^q(0)$ and are presented in unit of the model nuclear 
  magneton $\mu_N$.  \label{tab:1}}  
\end{table}

\begin{figure*}[ht]
  \begin{center}
\includegraphics[width=0.95\textwidth]{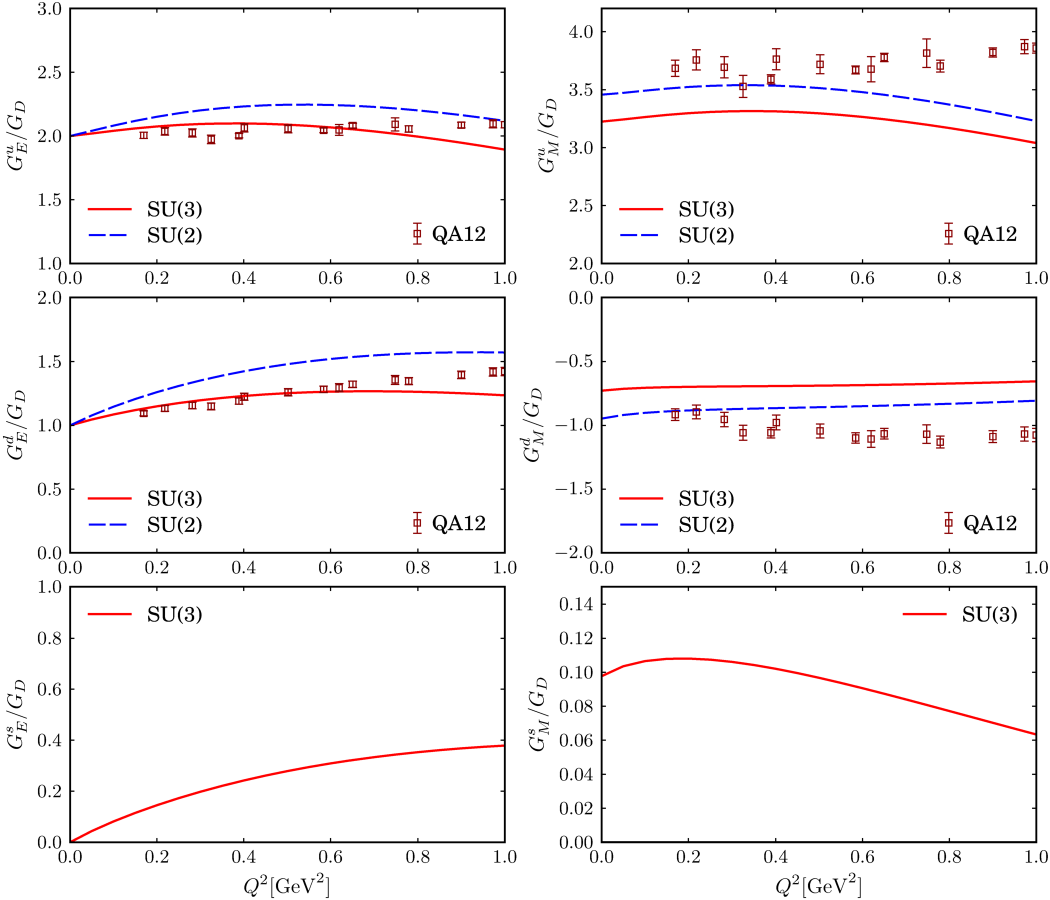}
  \end{center}
\caption{(Color online) Ratios of the nucleon Sachs flavour FFs to the dipole
  parameterizations (Eq.~(\ref{eq:GD}). The $u$ quark FFs in the upper 
  panel, the $d$ quark FFs in the middle panel, and the strange ones
  in the lower panel. The phenomenological data are taken from
  Refs.~\cite{Qattan:2012zf,Qattan:2012zf:Supplemental}
  (QA12). Notations are the same as in Fig.~\ref{fig:1}.}   
\label{fig:2}
\end{figure*} 
The Sachs FFs for the different quark flavours are presented
in Fig.~\ref{fig:2}, which are normalised by the dipole
parameterization defined as  
\begin{equation}
G_D(Q^2)=\frac{1}{\left(1+\frac{Q^2}{\Lambda_D^2} \right)^2}, \qquad
\Lambda_D^2=0.71\,\mathrm{GeV}^ 2 
\label{eq:GD}
\end{equation}
in comparison with the phenomenological data taken from
Refs.~\cite{Qattan:2012zf, Qattan:2012zf:Supplemental}, whose
normalisations at $Q^2=0$ are given as $G_M^u=3.67\,\mu_N$ and
$G_M^d=-1.03\,\mu_N$.  
The up and down electric FFs are more or less well reproduced. 
On the other hand, the up magnetic FF deviates from the data, as the
$Q^2$ increases. While the $Q^2$ dependence of the down magnetic FF
shows similar tendency to the data but the results seem a bit
overestimated. Since there are no corresponding experimental data yet
for the strange EMFFs, the lower panel of Fig.~\ref{fig:2} shows the 
predictions of the present model for the strange EMFFs. 

\section{Dirac and Pauli form factors}
The Dirac ($F_1$) and Pauli ($F_2$) FFs are expressed in terms of the
Sachs EMFFs inverting Eq.~(\ref{eq:DiracPauli2Sachs}), i.e.  
\begin{eqnarray}
F_1(Q^2)&=&\frac{G_E+\tau G_M}{1+\tau} \cr
F_2(Q^2)&=&\frac{G_M-G_E}{1+\tau}, 
\label{eq:Sachs2DiracPauli}
\end{eqnarray}
where $\tau$ is given by 
\begin{equation}
\tau=Q^2/(4 M^2).
\label{eq:tau}
\end{equation}

\begin{figure*}[ht]
\centerline{\includegraphics[width=0.49\textwidth]
{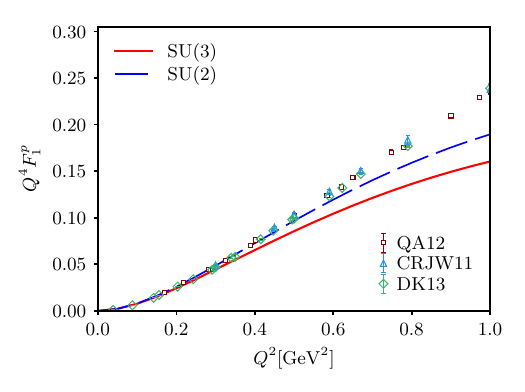} 
\includegraphics[width=0.49\textwidth]
{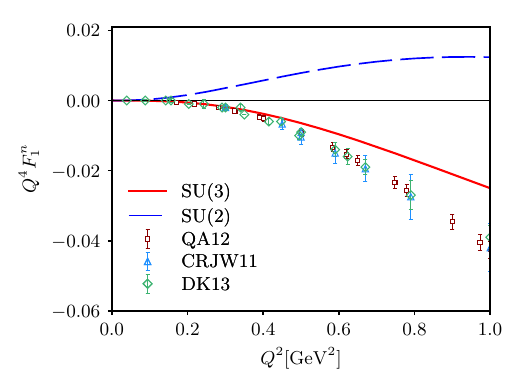}}\par
\centerline{\includegraphics[width=0.49\textwidth]
{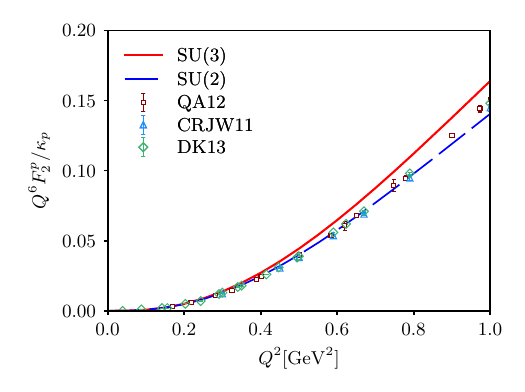} 
\includegraphics[width=0.49\textwidth]
{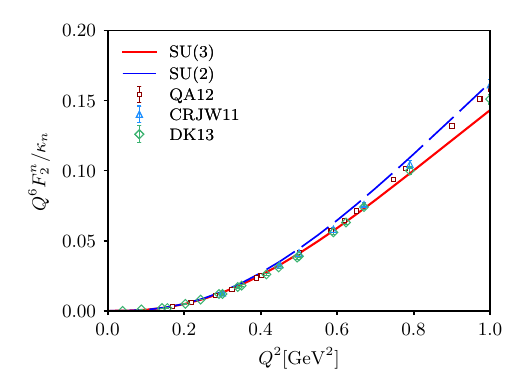}}
\caption{(Color online) Dirac FFs $F_1$ of the proton and the neutron, 
scaled with $Q^4$ in the upper panel and the Pauli FFs scaled with
$Q^6/\kappa_{p(n)}$ in the lower panel. The experimental data are
taken from Refs.~\cite{Cates:2011pz} (CRJW11),  
  \cite{Qattan:2012zf,Qattan:2012zf:Supplemental} (QA12)   and \cite{Diehl:2013xca} (DK13).  
Notations are the same as in Fig.~\ref{fig:1}.
}     
\label{fig:F1F2}
\end{figure*}
As mentioned in Introduction, pQCD with factorisation
schemes~\cite{Brodsky:1974vy} predicts that the nucleon Dirac FFs
scale with $1/Q^4$. It indicates that $Q^4 F_1(Q^2)$ becomes
asymptotically constant. Thus, the $Q^4 F_1(Q^2)$ is a more
interesting quantity than the $F_1$ itself. Figure~\ref{fig:F1F2}
shows the results for the nucleon Dirac FFs with $Q^4$ factor in
comparison with the experimental
data~\cite{Cates:2011pz,Qattan:2012zf,Diehl:2013xca,
  Qattan:2012zf:Supplemental}. The $Q^2$ dependence of 
$Q^4 F_1^p(Q^2)$ are well explained within the SU(2) model, while 
those from the corresponding SU(3) model seem slightly
underestimated, especially, as $Q^2$ increases. However, as for $Q^4
F_1^n(Q^2)$, the result of the SU(3) model describes the data well,
whereas the SU(2)   $F_1$ turns out positive.   As shown in the lower panel
of Fig.~\ref{fig:F1F2}, the results of $Q^6 F_2(Q^2)/\kappa$ both for 
the proton and the neutron are in good agreement with the experimental
data.   However, due to the momentum transfer range, the scaling behaviour is not clear.    

The flavour-decomposed Dirac ($F_1^q$) and Pauli ($F_2^q$) FFs 
are expressed as 
\begin{eqnarray}
F_{1,2}^u &=& 2 F_{1,2}^p+ F_{1,2}^n+F_{1,2}^s, \cr
F_{1,2}^d &=& F_{1,2}^p+2 F_{1,2}^n+F_{1,2}^s 
\label{eq:FqFNsu3}
\end{eqnarray}
in flavour SU(3). In flavour SU(2), the up and down Dirac and Pauli FFs
are simply written in terms of the corresponding proton and neutron
FFs. 
\begin{eqnarray}
F_{1,2}^u &=& 2 F_{1,2}^p+ F_{1,2}^n, \cr
F_{1,2}^d &=& F_{1,2}^p+2 F_{1,2}^n .
\label{eq:FqFNsu2}
\end{eqnarray}
  Note, however, that $F_{1,2}^{u,d}$ do not turn out the same in SU(3) and SU(2) 
just by neglecting $F_{1,2}^s$ since the flavour groups are different.   

\begin{figure*}[htb]  
\centerline{\includegraphics[width=0.49\textwidth]{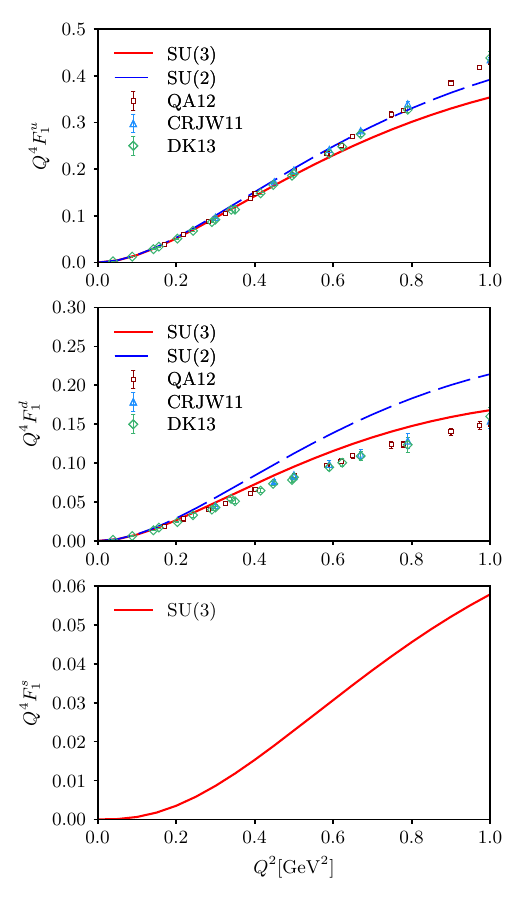}
\includegraphics[width=0.49\textwidth]{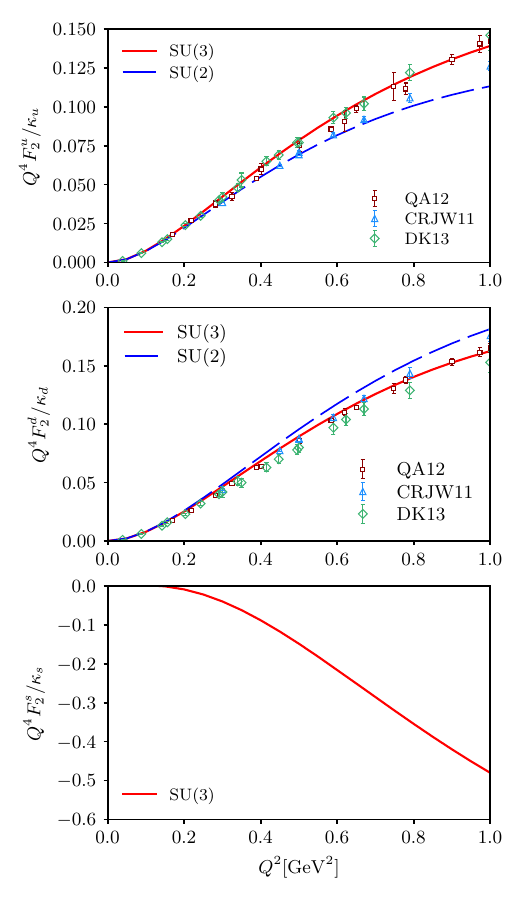}}
\caption{(Color online) The flavour-decomposed Dirac and Pauli FFs
  weighted by $Q^4$: The up FFs in the upper panel, the down ones in
  the middle panel, and the strange FFs in the lower panel. The
  experimental data are taken from Refs.~\cite{Cates:2011pz} (CRJW11),  
  \cite{Qattan:2012zf,Qattan:2012zf:Supplemental} (QA12) 
    and \cite{Diehl:2013xca} (DK13).   Notations are the same  as in Fig.~\ref{fig:1}.}  
\label{fig:flvQ4F}
\end{figure*} 
In Fig.~\ref{fig:flvQ4F}, we draw the results of $Q^4F_{1}^q$ and
$Q^4F_{2}^q/\kappa^q$ for the up ($u$), down ($d$) and strange ($s$)
quarks, respectively. $Q^4F_{1}^u$ shows stronger $Q^2$ dependence
than that of $Q^4F_{1}^d$ while $Q^4F_{2}^u$ exhibits weaker $Q^2$ 
dependence than that of $Q^4F_{2}^d$. The present results for both the
up and down quarks describe the data very well as in the case of the
proton and neutron FFs (see Fig.~\ref{fig:F1F2}). Again, we predict $Q^4
F_1^s$ and $Q^4F_2^s$. 

At $Q^ 2=0$ the Dirac and FFs are respectively
reduced to $F_1^p(0)= 1$, $F_1^n(0)=0$ and
$F_2^{p(n)}(0)=\kappa_{p(n)}$ with the corresponding anomalous
magnetic moment $\kappa_{p(n)}$ (See Table~\ref{tab:2}).  For the
flavour-decomposed Dirac FFs, with our normalization $F_1^u(0)= 2$, $F_1^d(0)= 1$ and 
$F_1^s(0)= 0$.   
\begin{table}[h] \begin{center}
\begin{tabular}{ccccccc} \hline
   & $\kappa_p$ & $\kappa_n$ & $\kappa_u$ & $\kappa_d$ & $\kappa_s$ \\
   \hline \hline 
SU(3) & $1.36$ & $-1.59$  & $1.22$ & $-1.73$ & $0.10$ \\
SU(2) & $1.62$ & $-1.78$ & $1.46$ & $-1.95$ &  \\
\mbox{Exp. \& Phen.} & $1.793$ & $-1.913$ &  $1.673$ & $-2.033$ &  \\ 
\hline
\end{tabular}
\end{center}
\caption{Anomalous magnetic moments $\kappa=F_2(0)$ for the proton and
  the nucleon. The flavour-decomposed anomalous magnetic moments are 
  also presented. Exp. \& Phen. denote the experimental data on the
  proton and the neutron anomalous magnetic moments, and the empirical
  data on the flavour-decomposed ones. 
\label{tab:2}}
\end{table}

\section{Transverse charge densities}
We are now in a position to discuss the quark transverse charge
densities inside both unpolarised and polarised nucleons. 
The traditional charge and magnetisation densities 
in the Breit framework are defined ambiguously 
because of the Lorentz contraction of the nucleon in its
moving direction~\cite{Kelly:2002if,Burkardt:2002hr}. To avoid this
ambiguity one can define the quark charge densities in the transverse
plane. Then, they provide essential information on how the 
charges and magnetisations of the quarks are distributed in the
transverse plane. 
When the nucleon is unpolarised, the quark transverse charge density
is defined as the two-dimensional Fourier transform of the nucleon
Dirac FFs  
\begin{equation}
\rho_{\rm ch}(b)= \frac{1}{(2\pi)^ 2} \int\! d^2 q\; e^{i{\bm q}\cdot{\bm b}}F_1 (Q^2)  
=\int_0^\infty \frac{dQ}{2\pi}\, Q J_0(Qb) F_1(Q^2) 
\label{eq:rhoch}
\end{equation}
where $b$ denotes the impact parameter, i.e. the distance in the
transverse plane to the place where the density is being probed, and
$J_0$ is a cylindrical Bessel Function of order 
zero~\cite{Miller:2007uy,Miller:2010nz}.  
Note that the Dirac FF at $Q^2=0$ and the anomalous magnetic moment
can be rederived from the transverse charge and magnetisation densities 
\begin{equation}
2\pi\! \int\! db\; b\, \rho_{\rm ch}(b) = F_1(0) ,\qquad
\pi\! \int\! db\; b\, \rho_{\rm m}(b)=\kappa, 
\label{eq:intrhoM}
\end{equation} 
either for the nucleon or for each individual   flavour, 
with the anomalous magnetisation density in the transverse
plane defined~\cite{Miller:2010nz,Venkat:2010by} by   
\begin{equation}
\rho_{\rm m} (b)= b \int_0^\infty \frac{dQ}{2\pi}\, Q^2 J_1(Qb) F_2(Q^2). 
\label{eq:rhoM}  
\end{equation}

By definition, Eq.s~(\ref{eq:rhoch},\ref{eq:rhoM}) seem to imply the 
knowledge of Dirac and Pauli form factors over a wide range of $Q^2$ 
in order to obtain meaningful densities.
This seems at odds with the fact that the present $\chi$CQSM FFs 
are obtained in the low transferred momenta.  
However, it turns out that with the model form factors, the integrals in 
Eq.s~(\ref{eq:rhoch},\ref{eq:rhoM}) are saturated in the range $Q^2<1.5$~(GeV/c)$^2$, \emph{i.e.} 
the computed densities do not change when the upper limit in the integrals
is set at different values above $1.5$~(GeV/c)$^2$.    

\begin{figure}[htb]
\centerline{
\includegraphics[width=0.49\textwidth]{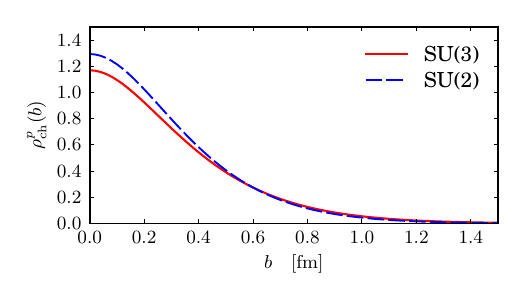}
\includegraphics[width=0.49\textwidth]{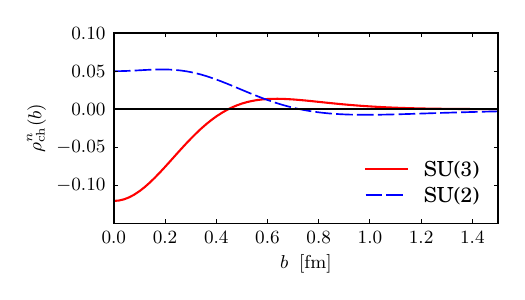}
}\par
\centerline{
\includegraphics[width=0.49\textwidth]{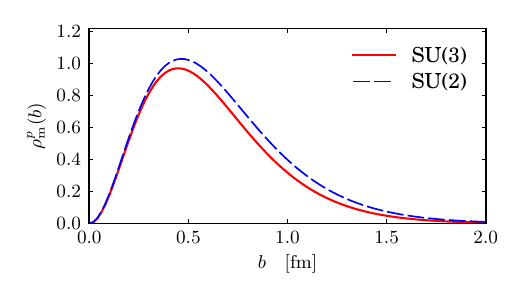}
\includegraphics[width=0.49\textwidth]{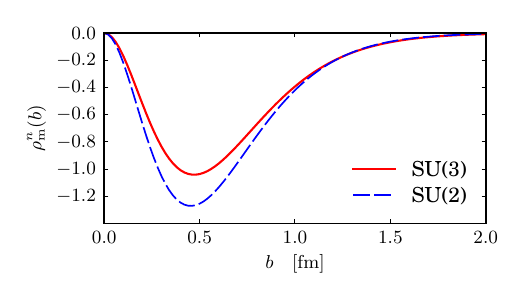}
}
\caption{(Color online) Transverse charge densities inside a proton
  (upper left panel) and a neutron (upper right panel), and the
  transverse magnetisation densities inside a proton (lower left
  panel) and a neutron (lower right panel). Notations are the same as
  in Fig.~\ref{fig:1}.} 
\label{fig:TCDs}  
\end{figure} 
In the upper panel of Fig.~\ref{fig:TCDs}, the transverse charge
densities inside both a proton and a neutron are drawn. The
results of the transverse charge density from the SU(2) model is
almost the same as that from the SU(3) model for the proton. However, 
it is very interesting to observe that the tranverse charge density
inside the neutron from the SU(2) model is opposite to that of the
SU(3) one. As already found in Figs.~\ref{fig:1} and \ref{fig:F1F2},
the SU(2) result is distinguished from the SU(3) one, mainly due to
the effects of the strange quark. These are in fact a surprising
results, because the SU(3) result interprets the inner structure of
the neutron totally differently from the SU(2) one: While the negative
charge, which mainly come from the down and strange quarks inside a
neutron, is centred on the neutron according to the SU(3) $\chi$QSM,
the SU(2) model suggests that the positive one be located in the
centre of the neutron. In Ref.~\cite{Miller:2007uy}, the transverse
charge density of the neutron was computed, based on the
parametrisation of the experimental EMFFs, and was found to be
negative in the centre of the neutron, which is in line with the
present result from the SU(3) model. To clarify this discrepancy
between the SU(2) and the SU(3) models, it might be essential to know
the strangeness content of the neutron. We will discuss later each
contribution of a quark with different flavour to the transverse
charge density inside the neutron more in detail. Another interesting
point in the transverse charge density inside a neutron is that it
turns positive as $b$ increases. The reason will soon be clear when we
discuss the flavour-decomposed transvese charge densities.

The lower panel of  Fig.~\ref{fig:TCDs} plots the transverse
magnetisation densities inside both a proton and a neutron. The
results from the SU(2) model are similar to those from the SU(3)
model.  As expected from their values of the anomalous magnetic
moments, the transverse magnetisation densities inside a proton are
positive but those inside a neutron turn out to be negative. We will
soon observe that the up quark and the down quark contribute 
oppositely to the magnetisation, which explains the results of the
transverse magnetisation densities inside a proton and a neutron.

\begin{figure}[htb]
\centerline{
\includegraphics[width=\textwidth]{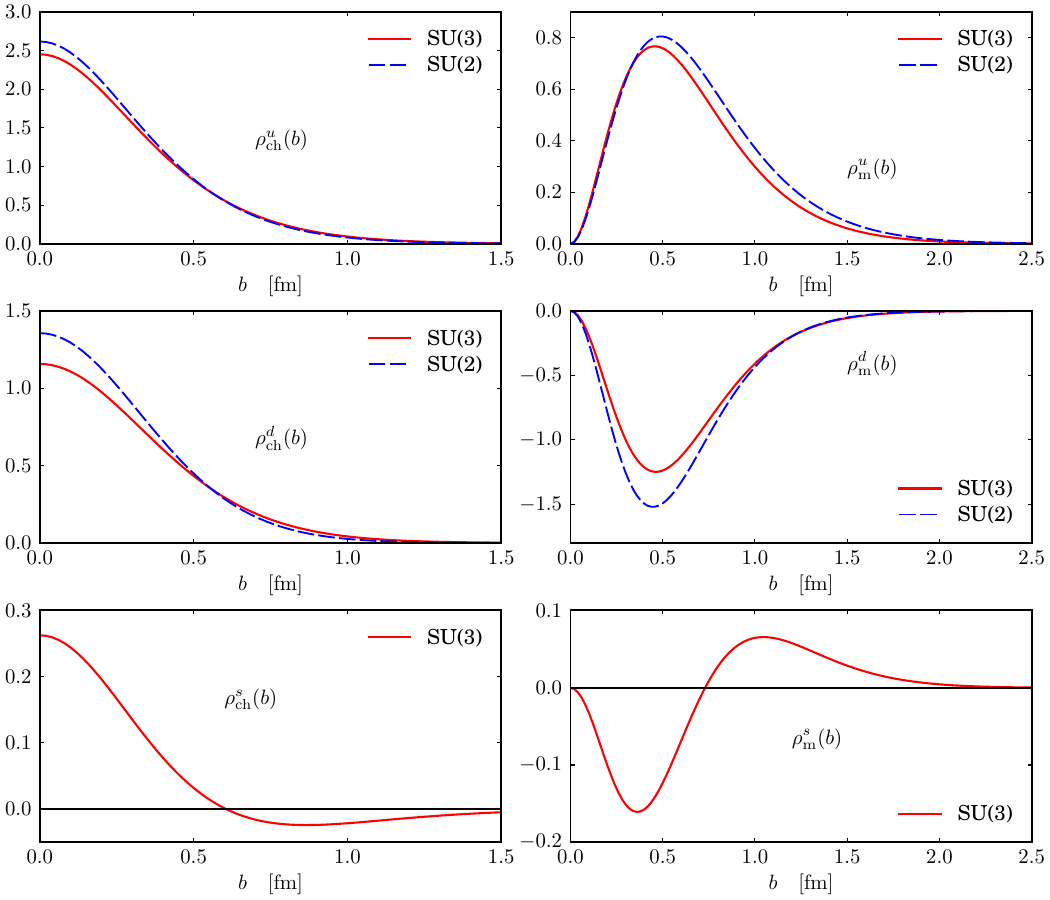}}

\caption{(Color online) Flavour-decomposed transverse charge and
  magnetisation densities inside a proton. Those for the up quark in
  the upper panel, the down ones in the middle panel, and the strange
  charge and magnetisation densities in the lower panel. Notations are
  the same as in Fig.~\ref{fig:1}. }  
\label{fig:flvTCDs}
\end{figure} 
In Fig.~\ref{fig:flvTCDs}, the transverse charge and magnetisation
densities are depicted for each flavour, in the left panel and the
right panel, respectively. The results of the transverse densities in
Fig.~\ref{fig:flvTCDs} do not include the charges for each flavor.
The charge densities for the up and the down quarks look similar to
the proton one shown in Fig.~\ref{fig:TCDs} and the SU(2) results
generally larger in the center region but fall off faster than the
SU(3) ones. The strange quark case shows interesting features. While
the charge density is found to be positive in the inner region, it  
becomes negative as $b$ increases. 
Note that the down quark inside a nucleon is more magnetised than the
up quark but was directed opposite to the up quark, which results in
the negative larger value of the anomalous magnetic moment for the
down quark than for the  up quark (see Tab.~\ref{tab:2}). The strange
transverse magnetisation densities look very different from those for
the up and down quarks: In the inner part of the nucleon, the strange
quark is negatively magnetised. As $b$ increases, the strange
magnetisation density turns positive. As a result, the strange
anomalous magnetic moment turns out to be small but positive:
$\kappa_s=0.10$ (see Tab.~\ref{tab:2}).     

As was discussed, the SU(3) transverse charge density was very
different from the SU(2) one. We can understand the reason for it from
the results of the flavour-decomposed transverse charge densities. 
The transverse charge densities inside a proton and a neutron can be
respectively expressed in terms of the flavour-decomposed ones 
\begin{eqnarray}
  \label{eq:pn}
 \rho_{\mathrm{ch}}^p &=& \frac13 (2\,\rho_{\mathrm{ch}}^u -
 \rho_{\mathrm{ch}}^d -\rho_{\mathrm{ch}}^s),\cr  
 \rho_{\mathrm{ch}}^n &=& \frac13 (2\,\rho_{\mathrm{ch}}^d -
 \rho_{\mathrm{ch}}^u - \rho_{\mathrm{ch}}^s). 
\end{eqnarray}
Since the $\rho_{\mathrm{ch}}^u$ governs  the transverse
charge density inside a proton ($\rho_{\mathrm{ch}}^p$) as shown
in Fig.~\ref{fig:flvTCDs}, the $\rho_{\mathrm{ch}}^s$ has almost  
negligible effects on it. However, when it comes to the
$\rho_{\mathrm{ch}}^n$, $2\,\rho_{\mathrm{ch}}^d$ and
$\rho_{\mathrm{ch}}^u$ in Eq.~(\ref{eq:pn}) are almost cancelled out
each other, which results in a small amount of the negative
density. In addition, $\rho_{\mathrm{ch}}^s$ contributes negatively to
the $\rho_{\mathrm{ch}}^n$, which finally leads to the negative value  
of the $\rho_{\mathrm{ch}}^n$ in the centred region, as shown in
Fig.~\ref{fig:TCDs}. In the case of the SU(2) model, the
$\rho_{\mathrm{ch}}^d$ turns out to be larger than that from the SU(3) 
model, so that the $\rho_{\mathrm{ch}}^n$ becomes positive but tiny. 
Thus, the strange transverse charge density, though it is small, plays
an essential role in explaining the negative value of the
$\rho_{\mathrm{ch}}^n$ in the centre of the neutron within the
framework of the $\chi$QSM. Moreover, the strange transverse charge 
density turns positive as $b$ increases. This explains partly
the reason why the $\rho_{\mathrm{ch}}^n$ becomes negative at higher
$b$. 

\begin{figure}[htb]
\centerline{\includegraphics[width=0.49\textwidth]
{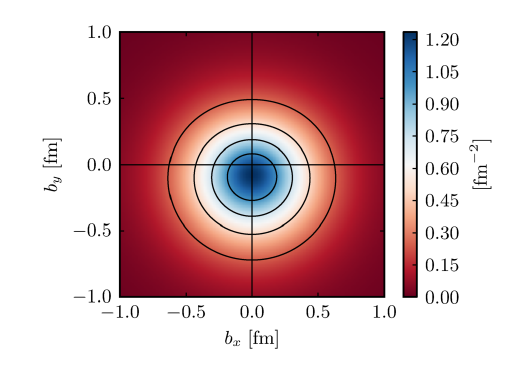}
\includegraphics[width=0.49\textwidth]
{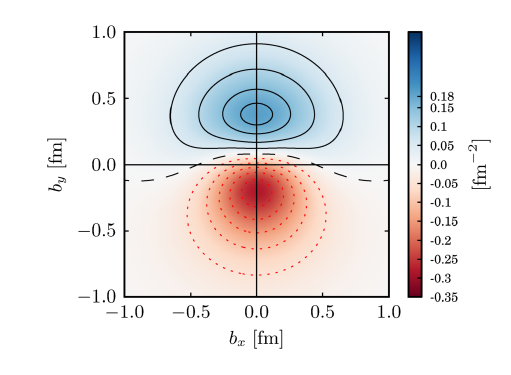}} \par 
\centerline{\includegraphics[width=0.49\textwidth]
{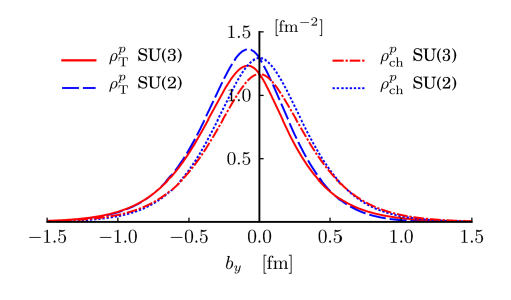}
\includegraphics[width=0.49\textwidth]
{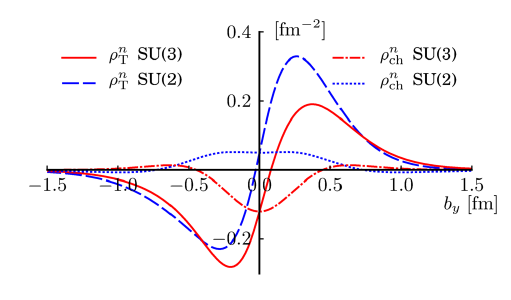}}

\caption{(Color online) Transverse charge densities inside a
  transversely polarised nucleon. The upper-left and upper-right
  panels show the transverse charge densities inside a proton and a
  neutron, respectively, being polarised along the $x$ axis. The lower
  panel depicts the corresponding transverse charge densities in the
  $y$ axis with $b_x$ fixed ($b_x=0$): The solid curve corresponds to
  the results of the transverse charge densities inside transversely
  polarised nucleons from the SU(3) model, whereas the dashed curve to
  those from the SU(2) model. The dotted and dash-dotted curves
  represent the SU(2) and SU(3) results for the transverse charge
  densities inside unpolarized nucleons, respectively. 
}
\label{fig:TCDpol}  
\end{figure} 
When the nucleon is transversely polarised along the $x$ axis, which 
can be described by the transverse spin operator of the nucleon $\bm
S_\perp =  \cos\phi_S \hat{\bm e}_x + \sin\phi_S \hat{\bm e}_y$,   
the transverse charge density inside a transversely polarised
nucleon is expressed~\cite{Carlson:2007xd} as
\begin{equation}
  \label{eq:trans_pol}
\rho_T (\bm b) \;=\; \rho_{\mathrm{ch}} (b) - \sin(\phi_b-\phi_S)
\frac{1}{2M_N b} \rho_{\mathrm{m}}(b),
\end{equation}
where $\rho_{\mathrm{m}}(b)$ is given in Eq.~(\ref{eq:rhoM}). The
position vector $\bm b$ from the centre of the nucleon in the 
transverse plane is denoted as $\bm b = b(\cos\phi_b \hat{\bm e}_x +
\sin \phi_b \hat{\bm e}_y)$. The $x$ axis is taken as the polarisation
direction of the nucleon, i.e. $\phi_S=0$.  
In the upper-left panel of Fig.~\ref{fig:TCDpol}, we plot the
transverse charge densities inside a transversely polarised proton. It
is shown that the charge density for the transversely polarised proton
is distorted in the negative $y$ direction. As discussed in
Refs.~\cite{Burkardt:2002hr,Carlson:2007xd}, the transverse
polarisation of the nucleon in the $x$ axis induces the electric
dipole moment along the negative $y$ direction, which is a well-known
relativistic effect. In the case of the neutron, the situation is even
more dramatic. As shown in the upper-right panel of
Fig.~\ref{fig:TCDs}, the negative charge is located at the centre of
the neutron with the positive charge surrounding it. However, when
the neutron is transversely polarised along the $x$ axis, the negative
charge is shifted to the negative $y$ direction but the positive one
is moved to the positive $y$ direction. This comes from the fact that
the neutron anomalous magnetic moment is negative, which yields an
induced electric dipole moment along the positive $y$ axis, as pointed
out by Ref.~\cite{Carlson:2007xd}.   

\begin{figure}[htb]
\centering{%
\includegraphics[width=\textwidth]{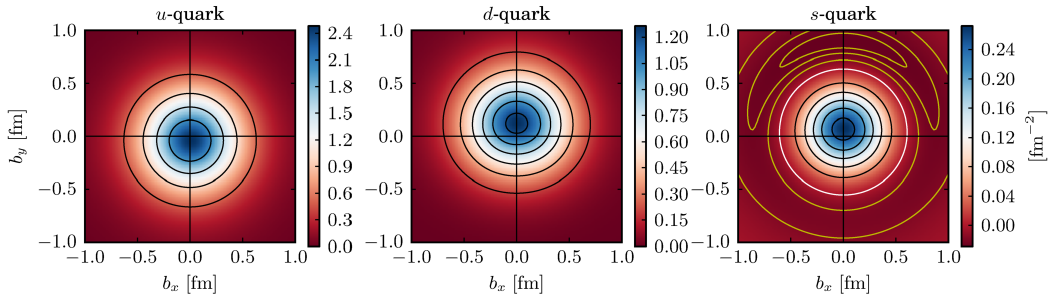}\\
\includegraphics[width=\textwidth]{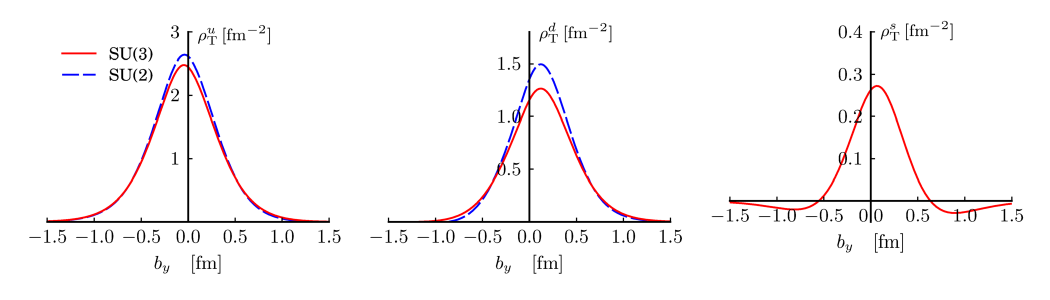}%
}
\caption{(Color online) Flavor-decomposed transverse charge densities
  inside a transversely polarised nucleon. The upper-left, the
  upper-middle and the upper-right panels draw the up, the down, and
  the strange transverse charge densities inside a proton and a neutron,
  respectively, being   polarised along the $x$ axis. The lower panel
  depicts the corresponding transverse charge densities in the $y$
  axis with $b_x$ fixed ($b_x=0$). Notations are the same as in
  Fig.~\ref{fig:1}.}     
\label{fig:flvTCDpol}  
\end{figure} 
It is very instructive to examine the transverse charge densities
inside the transversely polarised nucleon for each flavour, since they
reveal with more detail the inner structure of the nucleon. 
Figure~\ref{fig:flvTCDpol} illustrates them. The up transverse
charge density inside the transversely polarised nucleon, $\rho_T^u$
is shown to be shifted to the negative direction, while that for the
down quark is more distorted upwards. It is natural, since the up and
down quarks have positive and negative charges, respectively. However,
it is remarkable to see that the down quark is influenced more
strongly due to the transverse polarisation of the nucleon. The
$\rho_T^s$ is even more interesting. As discussed previously, the
strange anomalous magnetic moment is $\kappa_s=+0.10$, which would
induce the negative electric dipole moment along the negative
$b_y$. However, situation turns out to be more complicated. As shown
in the right panel of Fig.~\ref{fig:flvTCDpol}, the $\rho_{T}^s$ is
shifted to the positive $b_y$ and turns negative starting from
$b_y\approx 0.7\,\mathrm{fm}$. In order to understand this surprising
result, we need to reexamine the transverse magnetisation density for
the strange quark, which has been presented in
Fig.~\ref{fig:flvTCDs}. The strange magnetisation density is negative
in the inner part of the nucleon and then it becomes positive from  
$b\approx 0.7\,\mathrm{fm}$. Thus, the electric dipole moment is
correspondingly induced along the positive $y$ direction in the
centred region, and then it becomes negative from $b\approx
0.7\,\mathrm{fm}$, as drawn in the right panel of
Fig.~\ref{fig:flvTCDpol}.    

\section{Summary and conclusion}
In the present work, we aimed at investigating the electromagnetic
properties of the nucleon, based on the SU(2) and SU(3) chiral
quark-soliton model with a symmetry-preserving quantisation
employed. We considered the rotational $1/N_c$ corrections and the
first-order $m_s$ corrections.  
It should be stressed at this point that no free parameters were used in this work.
The only model parameter to be constrained in the baryon sector, namely 
the constituent quark mass, was taken from previous studies with various observables.   

We first presented the results of the ratio of the magnetic
form factor to the electric form factor of the proton. It was shown
that the results from the SU(2) chiral quark-soliton model described the
experimental data very well, whereas those of SU(3) seemed slightly
underestimated in higher $Q^2$. The general tendency of the present
results were in greement with the experimental data. As for the
neutron, the SU(3) results turned out to be rather different from
those in SU(2), which arose from the strange quark contribution to 
the neutron electric form factor. In particular, the neutron electric
form factor is rather sensitive to the tail of the soliton.  
We then discussed that the up and the down electric form factors
normalised by the dipole parametrisation were well reproduced in
comparison with the data. As for the magnetic form factors, they
deviate from the experimental data as $Q^2$ increases but the general 
behaviour of the form factors are in line with the experimental
data, which indicates that the $Q^2$ dependence are well explained. We
presented the prediction of the strange form factors normalised by the
dipole form factor.  

The Dirac and Pauli form factors were predicted to be asymptotically
proportional to $1/Q^4$ and $1/Q^6$ respectively in perturbative
QCD.   Thus, we studied $Q^4 F_1(Q^2)$ and $Q^6 F_2(Q^2)$ 
in order to compare their $Q^2$ dependence with the experimental data.   
We found that the present SU(2) model explained well $Q^4 F_1^p (Q^2)$ whereas the
result from the SU(3) model becomes underestimated at higher
$Q^2$. Both the SU(2) and SU(3) results for $Q^6
F_2^{p,n}(Q^2)/\kappa_{p,n}$ described the experimental data very
well. On the other hand, the SU(2) result for $Q^4 F_1^n$ is in
conflict with the data, but that from the SU(3) model is in agreement
with the data except for the higher $Q^2$ region. Again this
discrepancy can be understood by the sensitivity of the neutron
electric form factor to the soliton tail. The results for the
flavour-decomposed $Q^4 F_1(Q^2)$ and $Q^6 F_2(Q^2)$ were shown to be
generally in good agreement with the corresponding values from the experimental data.  

Having performed the two-dimensional Fourier transform of the nucleon
electromagnetic form factors, we were able to produce the charge
densities in the transverse plane inside a proton. As expected,
both the SU(2) and the SU(3) transverse charge densities were positive
in the proton. However, as for the neutron case, the result from the
SU(2) was opposite to that from the SU(3): the negative charge
was located in the centre of the neutron while the positive one was
distributed in outer part within the SU(3) chiral quark-soliton model,
it was other way around in the SU(2) model. The explanation comes from 
the decomposed-flavour transverse charge densities in the SU(3) model.  
In particular, the component of the strange quark played an essential
role in spite of the smallness of its magnitude. 
Since the up quark component mainly contributed to the 
transverse charge density inside a proton, the strange transverse
charge density was almost negligible. On the other hand, 
the up and the down quark contributions were nearly cancelled out in
such a way that the negative charge remained in the centre of the
neutron with small magnitude. Then the contribution of the strange
quark came into play, so that the transverse charge densities inside a
neutron finally became negative in the centre. 

When the proton was polarised along the positive $x$ direction, the
corresponding transverse charge density was shifted to the negative
$y$ direction, which indicated that the electric dipole moment was
induced along the negative $y$ direction. It is just a well-known
relativistic effect in electrodynamics. In the case of the neutron
polarised along the $x$ axis, the negative charge was moved to the
negative $y$ direction but the positive one was forced to the positive
$y$ axis. It implies that the neutron anomalous magnetic moment is
negative, which induces an electric dipole moment along the positive
$y$ axis. We also decomposed the transverse charge densities inside
the polarised nucleon for each flavour: the up transverse charge 
density for the nucleon transversely polarised along the positive $x$
axis was found to be shifted to the negative direction, while that of
the down quark was more distorted upwards. Since the up
and down quarks have positive and negative charges, respectively, one
can easily understand these features. However, the down quark was
found to be affected more strongly due to the transverse polarisation
of the nucleon. The strange charge density inside the transversely
polarised nucleon was shifted to the positive $b_y$ and turned out to
be negative in the outer region. This unexpected behavior of the 
strange charge density for the transversely polarised nucleon was
explained in terms of the strange magnetisation density. 

Since the transverse charge densities inside unpolarised and polarised
nucleons pave the novel way for understanding the internal structure
of the nucleon, it is interesting to investigate them for other
baryons such as the $\Delta$ isobar and hyperons. 
The transverse charge densities are directly connected to the
generalized parton distributions of which the integrations over parton
momentum fractions yield form factors and consequently the spatial
distribution of partons in the transverse plane. Moreover,  
the transverse charge densities for transition form factors provide a
new aspect of understanding the inner structure of the baryons. For
example, as Ref.~\cite{Carlson:2007xd} already studied, they exhibit
explicitly multipole structures of the transitions in the transverse
plane. Thus, it is of great importance to examine the transverse
charge densities for other baryons and for their transitions.   
Corresponding investigations are under way.

\section*{Acknowledgments}
H.-Ch.K. is grateful to R. Woloshyn and P. Navratil for their
hospitality during his stay at TRIUMF, where part of the present work
was done. The present work was supported by Basic Science 
Research Program through the National Research Foundation of Korea
funded by the Ministry of Education, Science and Technology (Grant
Number: NRF-2015R1D1A1A01060707).
  A. Silva is grateful to S.~Riordan for 
sharing information on experimental data on the nucleon eletromagnetic
form factors .


\vspace{1cm}


\begin{thebibliography}{99}

\bibitem{Jones:1999rz} 
  M.~K.~Jones {\it et al.}  [Jefferson Lab Hall A Collaboration],
  Phys.\ Rev.\ Lett.\  {\bf 84}, 1398 (2000)
  [nucl-ex/9910005].

\bibitem{Gayou:2001qt} 
  O.~Gayou {\it et al.}  [Jefferson Lab Hall A Collaboration], 
  Phys.\ Rev.\ C {\bf 64}, 038202 (2001).

\bibitem{Gayou:2001qd} 
  O.~Gayou {\it et al.}  [Jefferson Lab Hall A Collaboration],
  Phys.\ Rev.\ Lett.\  {\bf 88}, 092301 (2002)
  [nucl-ex/0111010].

\bibitem{Punjabi:2005wq} 
  V.~Punjabi {\it et al.}  [Jefferson Lab Hall A Collaboration], 
  Phys.\ Rev.\ C {\bf 71}, 055202 (2005)
  [Erratum-ibid.\ C {\bf 71}, 069902 (2005)]
  [nucl-ex/0501018].

\bibitem{Puckett:2010ac} 
  A.~J.~R.~Puckett, E.~J.~Brash, M.~K.~Jones, W.~Luo, M.~Meziane,
  L.~Pentchev, C.~F.~Perdrisat and V.~Punjabi {\it et al.}, 
  Phys.\ Rev.\ Lett.\  {\bf 104}, 242301 (2010)
  [arXiv:1005.3419 [nucl-ex]].

\bibitem{Bernauer:2010wm} 
  J.~C.~Bernauer {\it et al.}  [A1 Collaboration],
  Phys.\ Rev.\ Lett.\  {\bf 105}, 242001 (2010)
  [arXiv:1007.5076 [nucl-ex]].

\bibitem{Ron:2011rd} 
  G.~Ron {\it et al.}  [Jefferson Lab Hall A Collaboration],
  Phys.\ Rev.\ C {\bf 84}, 055204 (2011)
  [arXiv:1103.5784 [nucl-ex]].

\bibitem{Zhan:2011ji} 
  X.~Zhan, K.~Allada, D.~S.~Armstrong, J.~Arrington, W.~Bertozzi,
  W.~Boeglin, J.~-P.~Chen and K.~Chirapatpimol {\it et al.}, 
  Phys.\ Lett.\ B {\bf 705}, 59 (2011)
  [arXiv:1102.0318 [nucl-ex]].


\bibitem{HydeWright:2004gh} 
  C.~E.~Hyde-Wright and K.~de Jager,
  Ann.\ Rev.\ Nucl.\ Part.\ Sci.\  {\bf 54}, 217 (2004)
  [nucl-ex/0507001].

\bibitem{Arrington:2006zm} 
  J.~Arrington, C.~D.~Roberts and J.~M.~Zanotti,
  J.\ Phys.\ G {\bf 34}, S23 (2007)
  [nucl-th/0611050].

\bibitem{Perdrisat:2006hj} 
  C.~F.~Perdrisat, V.~Punjabi and M.~Vanderhaeghen,
  Prog.\ Part.\ Nucl.\ Phys.\  {\bf 59}, 694 (2007)
  [hep-ph/0612014].

\bibitem{Vanderhaeghen:2010nd} 
  M.~Vanderhaeghen and T.~Walcher,
  arXiv:1008.4225 [hep-ph].

\bibitem{Arrington:2011kb} 
  J.~Arrington, K.~de Jager and C.~F.~Perdrisat,
  J.\ Phys.\ Conf.\ Ser.\  {\bf 299}, 012002 (2011)
  [arXiv:1102.2463 [nucl-ex]].

\bibitem{Guichon:2003qm} 
  P.~A.~M.~Guichon and M.~Vanderhaeghen,
  Phys.\ Rev.\ Lett.\  {\bf 91}, 142303 (2003)
  [hep-ph/0306007].

\bibitem{Blunden:2003sp} 
  P.~G.~Blunden, W.~Melnitchouk and J.~A.~Tjon,
  Phys.\ Rev.\ Lett.\  {\bf 91}, 142304 (2003)
  [nucl-th/0306076].

\bibitem{Arrington:2004ae} 
  J.~Arrington,
  Phys.\ Rev.\ C {\bf 71}, 015202 (2005)
  [hep-ph/0408261].

\bibitem{Arrington:2007ux} 
  J.~Arrington, W.~Melnitchouk and J.~A.~Tjon,
  Phys.\ Rev.\ C {\bf 76}, 035205 (2007)
  [arXiv:0707.1861 [nucl-ex]].

\bibitem{Carlson:2007sp} 
  C.~E.~Carlson and M.~Vanderhaeghen,
  Ann.\ Rev.\ Nucl.\ Part.\ Sci.\  {\bf 57}, 171 (2007)
  [hep-ph/0701272 [HEP-PH]].

\bibitem{Arrington:2011dn} 
  J.~Arrington, P.~G.~Blunden and W.~Melnitchouk,
  Prog.\ Part.\ Nucl.\ Phys.\  {\bf 66}, 782 (2011)
  [arXiv:1105.0951 [nucl-th]].

\bibitem{Brodsky:1974vy} 
  S.~J.~Brodsky and G.~R.~Farrar,
  Phys.\ Rev.\ D {\bf 11}, 1309 (1975).

\bibitem{Aubert:2009mc} 
  B.~Aubert {\it et al.}  [BABAR Collaboration],
  Phys.\ Rev.\ D {\bf 80}, 052002 (2009)
  [arXiv:0905.4778 [hep-ex]].

\bibitem{Uehara:2012ag} 
  S.~Uehara {\it et al.}  [Belle Collaboration],
  Phys.\ Rev.\ D {\bf 86}, 092007 (2012)
  [arXiv:1205.3249 [hep-ex]].




\bibitem{Cates:2011pz} 
  G.~D.~Cates, C.~W.~de Jager, S.~Riordan and B.~Wojtsekhowski,
  Phys.\ Rev.\ Lett.\  {\bf 106}, 252003 (2011)
  [arXiv:1103.1808 [nucl-ex]].

\bibitem{Qattan:2012zf} 
  I.~A.~Qattan and J.~Arrington,
  Phys.\ Rev.\ C {\bf 86}, 065210 (2012)
  [arXiv:1209.0683 [nucl-ex]].

\bibitem{Qattan:2015qxa} 
  I.~A.~Qattan, J.~Arrington and A.~Alsaad,
  Phys.\ Rev.\ C {\bf 91}, no. 6, 065203 (2015)
  [arXiv:1502.02872 [nucl-ex]].

 \bibitem{Diehl:2013xca} 
  M.~Diehl and P.~Kroll,
  Eur.\ Phys.\ J.\ C {\bf 73}, no. 4, 2397 (2013)
  [arXiv:1302.4604 [hep-ph]].  

 \bibitem{Cloet:2008re} 
  I.~C.~Cloet, G.~Eichmann, B.~El-Bennich, T.~Klahn and C.~D.~Roberts,
  Few Body Syst.\  {\bf 46}, 1 (2009)
  [arXiv:0812.0416 [nucl-th]].  

 \bibitem{Crawford:2010gv} 
  C.~Crawford {\it et al.},
  Phys.\ Rev.\ C {\bf 82}, 045211 (2010)
  [arXiv:1003.0903 [nucl-th]].  

 \bibitem{Riordan:2010id} 
  S.~Riordan {\it et al.},
  Phys.\ Rev.\ Lett.\  {\bf 105}, 262302 (2010)
  [arXiv:1008.1738 [nucl-ex]].  

\bibitem{Eichmann:2011vu} 
  G.~Eichmann,
  Phys.\ Rev.\ D {\bf 84}, 014014 (2011)
  [arXiv:1104.4505 [hep-ph]].

\bibitem{Rohrmoser:2011tw} 
  M.~Rohrmoser, K.~-S.~Choi and W.~Plessas,
  arXiv:1110.3665 [hep-ph].

 \bibitem{Rohrmoser:2017zpk} 
  M.~Rohrmoser, K.~S.~Choi and W.~Plessas,
  Few Body Syst.\  {\bf 58}, no. 2, 83 (2017)
  doi:10.1007/s00601-017-1243-0
  [arXiv:1701.07337 [hep-ph]].     

\bibitem{Cloet:2012cy} 
  I.~C.~Cloet and G.~A.~Miller,
  Phys.\ Rev.\ C {\bf 86}, 015208 (2012)
  [arXiv:1204.4422 [nucl-th]].

\bibitem{GonzalezHernandez:2012jv} 
  J.~O.~Gonzalez-Hernandez, S.~Liuti, G.~R.~Goldstein and K.~Kathuria,
  Phys.\ Rev.\ C {\bf 88}, no. 6, 065206 (2013)
  [arXiv:1206.1876 [hep-ph]].

 \bibitem{Chakrabarti:2013dda} 
  D.~Chakrabarti and C.~Mondal,
  Eur.\ Phys.\ J.\ C {\bf 73}, 2671 (2013)
  [arXiv:1307.7995 [hep-ph]].     

 \bibitem{Mondal:2015uha} 
  C.~Mondal and D.~Chakrabarti,
  Eur.\ Phys.\ J.\ C {\bf 75}, no. 6, 261 (2015)
  [arXiv:1501.05489 [hep-ph]].   

 \bibitem{Mondal:2016xpk} 
  C.~Mondal,
  Phys.\ Rev.\ D {\bf 94}, no. 7, 073001 (2016)
  [arXiv:1609.07759 [hep-ph]].   

 \bibitem{Sharma:2016cnf} 
  N.~Sharma,
  Eur.\ Phys.\ J.\ A {\bf 52}, no. 11, 338 (2016)
  [arXiv:1610.07745 [hep-ph]].     

 \bibitem{Nikkhoo:2015jzi} 
  N.~S.~Nikkhoo and M.~R.~Shojaei,
  Int.\ J.\ Mod.\ Phys.\ E {\bf 24}, no. 11, 1550086 (2015).  
    
 \bibitem{Brodsky:2016uln} 
  S.~J.~Brodsky, R.~F.~Lebed and V.~E.~Lyubovitskij,
  Phys.\ Lett.\ B {\bf 764}, 174 (2017)
  [arXiv:1609.06635 [hep-ph]].     

 \bibitem{Obukhovsky:2014xja} 
  I.~T.~Obukhovsky, A.~Faessler, T.~Gutsche and V.~E.~Lyubovitskij,
  J.\ Phys.\ G {\bf 41}, 095005 (2014).   
  
 \bibitem{Chakrabarti:2014dna} 
  D.~Chakrabarti and C.~Mondal,
  Eur.\ Phys.\ J.\ C {\bf 74}, 2962 (2014)
  [arXiv:1402.4972 [hep-ph]].   
  
 \bibitem{Sharma:2014voa} 
  N.~Sharma,
  Phys.\ Rev.\ D {\bf 90}, no. 9, 095024 (2014)
  [arXiv:1411.7486 [hep-ph]].   
  
\bibitem{Diakonov:1987ty} 
  D.~Diakonov, V.~Y.~Petrov and P.~V.~Pobylitsa,
  Nucl.\ Phys.\ B {\bf 306}, 809 (1988).

\bibitem{Wakamatsu:1990ud} 
  M.~Wakamatsu and H.~Yoshiki,
  Nucl.\ Phys.\ A {\bf 524}, 561 (1991).


%

\bibitem{Blotz:1992pw} 
  A.~Blotz, D.~Diakonov, K.~Goeke, N.~W.~Park, V.~Y.~Petrov and
  P.~V.~Pobylitsa, 
  Nucl.\ Phys.\ A {\bf 555}, 765 (1993).

\bibitem{Alkofer:1994ph}
R. Alkofer, H. Reinhardt and H. Weigel,
Phys. Rept. {\bf 265}, 139 (1996).

\bibitem{Christov:1995vm} 
  C.~V.~Christov, A.~Blotz, H.~-Ch.~Kim, P.~Pobylitsa, T.~Watabe,
  T.~Meissner, E.~Ruiz Arriola and K.~Goeke, 
  Prog.\ Part.\ Nucl.\ Phys.\  {\bf 37}, 91 (1996)
  [hep-ph/9604441].

\bibitem{Weigel:2008zz} 
  H.~Weigel,
  Lect.\ Notes Phys.\  {\bf 743}, 1 (2008).

\bibitem{Goeke:2001tz} 
  K.~Goeke, M.~V.~Polyakov and M.~Vanderhaeghen,
  Prog.\ Part.\ Nucl.\ Phys.\  {\bf 47}, 401 (2001)
  [hep-ph/0106012].

\bibitem{Goeke:2006gi} 
  K.~Goeke, H.~-Ch.~Kim, A.~Silva and D.~Urbano,
  Eur.\ Phys.\ J.\ A {\bf 32}, 393 (2007)
  [hep-ph/0608262].

\bibitem{Silva:2005qm} 
  A.~Silva, H.~-Ch.~Kim, D.~Urbano and K.~Goeke,
  Phys.\ Rev.\ D {\bf 74}, 054011 (2006)
  [hep-ph/0601239].

\bibitem{Kim:1995mr} 
  H.~-Ch.~Kim, A.~Blotz, M.~V.~Polyakov and K.~Goeke,
  Phys.\ Rev.\ D {\bf 53}, 4013 (1996)
  [hep-ph/9504363].

\bibitem{Praszalowicz:1998jm} 
  M.~Prasza\l owicz, T.~Watabe and K.~Goeke,
  Nucl.\ Phys.\ A {\bf 647}, 49 (1999)
  [hep-ph/9806431].

\bibitem{Witten:1979kh} 
  E.~Witten,
  Nucl.\ Phys.\ B {\bf 160}, 57 (1979) and
\bibitem{Witten:1983} E.~Witten,
  Nucl.\ Phys.\ B {\bf 223}, 422 (1983)  and 
  Nucl.\ Phys.\ B {\bf 223}, 433 (1983).

\bibitem{Yang:2016qdz} 
  G.~S.~Yang, H.-Ch.~Kim, M.~V.~Polyakov and M.~Prasza{\l}wicz,
  Phys.\ Rev.\ D {\bf 94}, 071502 (2016)
  doi:10.1103/PhysRevD.94.071502
  [arXiv:1607.07089 [hep-ph]].

\bibitem{Kim:2017jpx} 
  H.-Ch.~Kim, M.~V.~Polyakov and M.~Prasza{\l}owicz,
  Phys.\ Rev.\ D {\bf 96}, no. 1, 014009 (2017)
  Addendum: [Phys.\ Rev.\ D {\bf 96}, no. 3, 039902 (2017)]
  doi:10.1103/PhysRevD.96.039902, 10.1103/PhysRevD.96.014009
  [arXiv:1704.04082 [hep-ph]].

\bibitem{Kim:2017khv} 
  H.-Ch.~Kim, M.~V.~Polyakov, M.~Praszalowicz and G.~S.~Yang,
  Phys.\ Rev.\ D {\bf 96}, no. 9, 094021 (2017)
  doi:10.1103/PhysRevD.96.094021
  [arXiv:1709.04927 [hep-ph]].

\bibitem{Androic:2009aa} 
  D.~Androic {\it et al.}  [G0 Collaboration],
  Phys.\ Rev.\ Lett.\  {\bf 104}, 012001 (2010)
  [arXiv:0909.5107 [nucl-ex]].

\bibitem{Ji:1998pc} 
  X.~-D.~Ji,
  J.\ Phys.\ G {\bf 24}, 1181 (1998)
  [hep-ph/9807358].

\bibitem{Diehl:2003ny} 
  M.~Diehl,
  Phys.\ Rept.\  {\bf 388}, 41 (2003)
  [hep-ph/0307382].

\bibitem{Belitsky:2005qn} 
  A.~V.~Belitsky and A.~V.~Radyushkin,
  Phys.\ Rept.\  {\bf 418}, 1 (2005)
  [hep-ph/0504030].
  
\bibitem{Ernst:1960zza} 
  F.~J.~Ernst, R.~G.~Sachs, and K.~C.~Wali,
  Phys.\ Rev.\  {\bf 119}, 1105 (1960).
  
\bibitem{Sachs:1962zzc} 
  R.~G.~Sachs,
  Phys.\ Rev.\  {\bf 126}, 2256 (1962).
   
\bibitem{Arnold:1980zj} 
  R.~G.~Arnold, C.~E.~Carlson, and F.~Gross,
  Phys.\ Rev.\ C {\bf 23}, 363 (1981).

\bibitem{Miller:2007uy} 
  G.~A.~Miller,
  Phys.\ Rev.\ Lett.\  {\bf 99}, 112001 (2007)
  [arXiv:0705.2409 [nucl-th]].

\bibitem{Miller:2010nz} 
  G.~A.~Miller,
  Ann.\ Rev.\ Nucl.\ Part.\ Sci.\  {\bf 60}, 1 (2010)
  [arXiv:1002.0355 [nucl-th]].

\bibitem{Venkat:2010by} 
  S.~Venkat, J.~Arrington, G.~A.~Miller and X.~Zhan,
  Phys.\ Rev.\ C {\bf 83}, 015203 (2011)
  [arXiv:1010.3629 [nucl-th]].

\bibitem{Milbrath:1997de}
  B.~D.~Milbrath {\it et al.}  [Bates FPP Collaboration],
  Phys.\ Rev.\ Lett.\  {\bf 80} (1998) 452
   [Erratum-ibid.\  {\bf 82} (1999) 2221]
  [nucl-ex/9712006].
  
\bibitem{Pospischil:2001pp} 
  T.~Pospischil {\it et al.}  [A1 Collaboration],
  Eur.\ Phys.\ J.\ A {\bf 12}, 125 (2001).

\bibitem{MacLachlan:2006vw} 
  G.~MacLachlan, A.~Aghalarian, A.~Ahmidouch, B.~D.~Anderson, R.~Asaturian, O.~Baker, A.~R.~Baldwin and D.~Barkhuff {\it et al.},
  Nucl.\ Phys.\ A {\bf 764}, 261 (2006).

\bibitem{Paolone:2010qc} 
  M.~Paolone, S.~P.~Malace, S.~Strauch, I.~Albayrak, J.~Arrington, B.~L.~Berman, E.~J.~Brash and W.~J.~Briscoe {\it et al.},
  Phys.\ Rev.\ Lett.\  {\bf 105}, 072001 (2010)
  [arXiv:1002.2188 [nucl-ex]].
  
\bibitem{Puckett:2011xg} 
  A.~J.~R.~Puckett, E.~J.~Brash, O.~Gayou, M.~K.~Jones, L.~Pentchev, C.~F.~Perdrisat, V.~Punjabi and K.~A.~Aniol {\it et al.},
  Phys.\ Rev.\ C {\bf 85}, 045203 (2012)
  [arXiv:1102.5737 [nucl-ex]].
  
\bibitem{Jones:2006kf} 
  M.~K.~Jones {\it et al.}  [Resonance Spin Structure Collaboration],
  Phys.\ Rev.\ C {\bf 74}, 035201 (2006)
  [nucl-ex/0606015].
  
\bibitem{Crawford:2006rz} 
  C.~B.~Crawford, A.~Sindile, T.~Akdogan, R.~Alarcon, W.~Bertozzi, E.~Booth, T.~Botto and J.~Calarco {\it et al.},
  Phys.\ Rev.\ Lett.\  {\bf 98}, 052301 (2007)
  [nucl-ex/0609007].
 
\bibitem{Passchier:1999cj} 
  I.~Passchier, R.~Alarcon, T.~S.~Bauer, D.~Boersma, J.~F.~J.~van den
  Brand, L.~D.~van Buuren, H.~J.~Bulten and M.~Ferro-Luzzi {\it et al.}, 
 Phys.\ Rev.\ Lett.\  {\bf 82}, 4988 (1999)
 [nucl-ex/9907012].

\bibitem{Zhu:2001md} 
  H.~Zhu {\it et al.}  [Jefferson Lab E93-026 Collaboration],
  Phys.\ Rev.\ Lett.\  {\bf 87}, 081801 (2001)
  [nucl-ex/0105001].

\bibitem{Warren:2003ma} 
  G.~Warren {\it et al.}  [Jefferson Lab E93-026 Collaboration],
  Phys.\ Rev.\ Lett.\  {\bf 92}, 042301 (2004)
  [nucl-ex/0308021].

\bibitem{Geis:2008aa} 
  E.~Geis {\it et al.}  [BLAST Collaboration],
  Phys.\ Rev.\ Lett.\  {\bf 101}, 042501 (2008)
  [arXiv:0803.3827 [nucl-ex]].

\bibitem{Herberg:1999ud} 
  C.~Herberg, M.~Ostrick, H.~G.~Andresen, J.~R.~M.~Annand,
  K.~Aulenbacher, J.~Becker, P.~Drescher and D.~Eyl {\it et al.}, 
  Eur.\ Phys.\ J.\ A {\bf 5}, 131 (1999).

\bibitem{Glazier:2004ny} 
  D.~I.~Glazier {\it et al.} [A1 Collaboration], 
  Eur.\ Phys.\ J.\ A {\bf 24}, 101 (2005)
  [nucl-ex/0410026].

\bibitem{Plaster:2005cx} 
  B.~Plaster {\it et al.}  [Jefferson Lab E93-038 Collaboration],
  Phys.\ Rev.\ C {\bf 73}, 025205 (2006)
  [nucl-ex/0511025].

\bibitem{Bermuth:2003qh} 
  J.~Bermuth, P.~Merle, C.~Carasco, D.~Baumann, D.~Bohm, D.~Bosnar,
  M.~Ding and M.~O.~Distler {\it et al.}, 
  Phys.\ Lett.\ B {\bf 564}, 199 (2003)
  [nucl-ex/0303015].

\bibitem{Silva:2001st} 
  A.~Silva, H.~-Ch.~Kim and K.~Goeke,
  Phys.\ Rev.\ D {\bf 65}, 014016 (2002)
  [Erratum-ibid.\ D {\bf 66}, 039902 (2002)]
  [hep-ph/0107185].

\bibitem{Qattan:2012zf:Supplemental} 
  Supplemental Material from \cite{Qattan:2012zf} 
at  http://link.aps.org/supplemental/10.1103/PhysRevC.86.065210

\bibitem{Kelly:2002if}
J.~J.~Kelly,
  Phys.\ Rev.\ C {\bf 66}, 065203 (2002)
  [hep-ph/0204239].
 
\bibitem{Burkardt:2002hr}
  M.~Burkardt,
  Int.\ J.\ Mod.\ Phys.\  {\bf A18 } (2003)  173-208.
  [arXiv:hep-ph/0207047 [hep-ph]].


\bibitem{Carlson:2007xd} 
  C.~E.~Carlson and M.~Vanderhaeghen,
  Phys.\ Rev.\ Lett.\  {\bf 100}, 032004 (2008)
  [arXiv:0710.0835 [hep-ph]].

\end{thebibliography}
\end{document}